\documentclass[manuscript,screen]{acmart}

\usepackage{subcaption}
\usepackage[utf8]{inputenc}
\usepackage{bm}
\usepackage{comment}
\usepackage{titlesec}
\usepackage{array}
\usepackage[most]{tcolorbox}
\PassOptionsToPackage{table}{xcolor}
\usepackage{xcolor}
\usepackage{geometry}
\usepackage{pifont}
\usepackage{tabularx}
\usepackage{tikz}
\usepackage{textcomp}
\usepackage{makecell}
\usepackage{xspace}
\usepackage{amsmath}
\usepackage{amssymb}
\usepackage{enumitem}
\usepackage{booktabs}
\usepackage{ragged2e}
\usepackage{rotating}
\usepackage{fontawesome5}
\usepackage{caption}
\usepackage{colortbl}
\usepackage{multirow}
\usepackage{graphicx}
\usepackage{mdframed}
\usepackage{needspace}
\usepackage[table]{xcolor}
\usepackage{colortbl}
\usepackage{xcolor}
\usepackage{multirow}
\usepackage{booktabs}
\usepackage{wrapfig}

\definecolor{lightgray}{gray}{0.5}
\definecolor{medgray}{gray}{1}
\definecolor{highlight}{HTML}{FFD700}
\definecolor{rowgray}{HTML}{c2c3c4} 

\newcolumntype{M}{>{\columncolor{ColBG}}c}
\newcolumntype{L}{>{\raggedright\arraybackslash}l}

\newcommand{\linebreakand}{%
  \end{@IEEEauthorhalign}
  \hfill\mbox{}\par
  \mbox{}\hfill\begin{@IEEEauthorhalign}
}

\usepackage{amsmath,amsfonts}
\usepackage{caption}
\usepackage{blindtext}
\usepackage{tcolorbox}
\usepackage[final]{pdfpages}
\usepackage{lipsum,multicol}
\usepackage{xcolor}
\usepackage{tikz}
\usepackage{listings}
\usepackage{enumitem}
\usepackage{hyperref}
\usepackage{amsfonts}
\usepackage{wrapfig}
\usepackage{subcaption} 
\usepackage{adjustbox}
\usepackage{colortbl}
\usepackage{fancybox}
\usepackage{multirow}
\usepackage[normalem]{ulem}
\useunder{\uline}{\ul}{}
\usepackage{graphicx}
\usepackage{booktabs}
\usepackage{svg}
\usepackage[utf8]{inputenc}

\definecolor{highlightyellow}{RGB}{255, 255, 153}
\definecolor{featuregreen1}{RGB}{198, 239, 206}
\definecolor{featuregreen2}{RGB}{220, 250, 220}
\definecolor{featuregreen3}{RGB}{255, 255, 204}
\definecolor{featuregreen4}{RGB}{255, 250, 190}
\definecolor{featuregray}{gray}{0.95}

\lstdefinestyle{mystyle}{
  language=Python,
  basicstyle=\fontsize{6}{7}\selectfont\ttfamily,
  keywordstyle=\color{blue},
  commentstyle=\color{gray},
  stringstyle=\color{red},
  showstringspaces=false,
  breaklines=true,
  frame=single,
  backgroundcolor=\color{featuregray}
}

\newcommand{\approach}{{\texttt{FeatureSHAP}}\xspace}
\newcommand{\nparticipants}{{37}\xspace}

\newcommand{\ie}{\textit{i.e.,}\xspace}
\newcommand{\eg}{\textit{e.g.,}\xspace}
\newcommand{\etc}{\textit{etc.}\xspace}
\newcommand{\etal}{et al.\xspace}

\definecolor{darkgreen}{rgb}{0.0, 0.5, 0.0}

\makeatletter
\newcommand*{\radiobutton}{%
  \@ifstar{\@radiobutton0}{\@radiobutton1}%
}
\newcommand*{\@radiobutton}[1]{%
  \begin{tikzpicture}[baseline={(0,-0.6ex)}]
    \pgfmathsetlengthmacro\radius{height("X")/2}
    \draw[radius=\radius] circle;
    \ifcase#1 \fill[radius=.6*\radius] circle;\fi
  \end{tikzpicture}%
}
\makeatother

\newcommand{\RQ}[1]{RQ\textsubscript{#1}\xspace}

\definecolor{Gray}{gray}{0.9}
\definecolor{codegreen}{rgb}{0,0.6,0}
\definecolor{codegray}{rgb}{0.73,0.38,0.06}
\definecolor{codepurple}{rgb}{0.70,0.27,0}
\definecolor{codemagenta}{rgb}{0.74,0.09,0.42}
\definecolor{codeoutput}{rgb}{0.5,0,0}
\definecolor{backcolour}{rgb}{0.96,0.96,0.96}

\definecolor{gray50}{gray}{.5}
\definecolor{gray40}{gray}{.6}
\definecolor{gray30}{gray}{.7}
\definecolor{gray20}{gray}{.8}
\definecolor{gray10}{gray}{.9}
\definecolor{gray05}{gray}{.95}

\newenvironment{examplebox}{\par\begingroup
  \setlength{\fboxsep}{5pt}
  \hspace{-0.4cm}
  \setbox0=\vbox\bgroup\noindent
  \hsize=0.95\linewidth
  \begin{minipage}{0.95\linewidth}\normalsize}
  {\end{minipage}\egroup
  \textcolor{gray20}{\fboxsep1.5pt\fbox
    {\fboxsep5pt\colorbox{gray05}{\normalcolor\box0}}}
  \endgroup\par\noindent
  \normalcolor\ignorespacesafterend}

\newtcolorbox{promptbox}{colback=white, arc=0.5mm, top=1mm, bottom=1mm, left=1mm, right=1mm, title=System prompt used for generation}

\newcommand{\mycolorbox}[2]{%
\begin{center}
\begin{tcolorbox}[
  enhanced,
  width=0.93\linewidth,
  colback=white,
  colframe=black,
  boxrule=0.2mm,
  arc=5pt,
  left=10pt, right=10pt, top=7pt, bottom=6pt,
  fonttitle=\bfseries,
  coltitle=white,
  title={\textbf{Answer to \RQ{#1}.}},
  attach boxed title to top left={
      yshift=-6pt,
      xshift=-3pt
  },
  boxed title style={
      colback=black,
      colframe=black,
      arc=5pt,
      top=0.5pt,
      bottom=1pt,
      left=6pt,
      right=6pt
  }
]
#2
\end{tcolorbox}
\end{center}
}

\newtcolorbox{boxK}{
    fontupper = \small,
    sharpish corners, % better drop shadow
    boxrule = 0pt,
    toprule = 0pt, % top rule weight
}

\newcommand{\secref}[1]{Sec.~\ref{#1}\xspace}
\newcommand{\figref}[1]{Fig.~\ref{#1}\xspace}
\newcommand{\tabref}[1]{Tab.~\ref{#1}\xspace}

\tcbuselibrary{skins, breakable}

\definecolor{lightgray}{rgb}{0.95, 0.95, 0.95}
\definecolor{darkgray}{rgb}{0.7, 0.7, 0.7}
\definecolor{mygreen}{rgb}{0.0, 0.5, 0.0}
\definecolor{myred}{rgb}{0.8, 0.0, 0.0}

\newcolumntype{C}{>{\bfseries}c}
\definecolor{main}{HTML}{cccccc}
\definecolor{sub}{HTML}{000000}

\newtcolorbox{boxM}{
    fontupper = \color{blue},
    rounded corners,
    arc = 6pt,
    colback = main!80, 
    colframe = main, 
    boxrule = 0pt, 
    bottomrule = 4.5pt,
    enhanced,
    fuzzy shadow = {0pt}{-3pt}{-0.5pt}{0.5pt}{blue!35},
    overlay={
        \node[anchor=north east] at (frame.north east) {
            \includegraphics[width=15pt]{images/future.png}
        };
    }
}

\AtBeginDocument{%
  }

\begin{document}

\title{Toward Explaining Large Language Models in Software Engineering Tasks}

\author{Antonio Vitale}
\authornote{Both authors contributed equally to this research.}
\affiliation{%
  \institution{University of Molise \& Politecnico di Torino}
  \city{Termoli}
  \country{Italy}
}
\email{antonio.vitale@unimol.it}\email{antonio.vitale@polito.it}

\author{Khai Nguyen-Nguyen}
\authornotemark[1]
\affiliation{%
  \institution{William \& Mary}
  \city{Williamsburg}
  \state{Virginia}
  \country{USA}
}
\email{knguyen07@wm.edu}

\author{Denys Poshyvanyk}
\email{dposhyvanyk@wm.edu}
\affiliation{%
  \institution{William \& Mary}
  \city{Williamsburg}
  \state{VA}
  \country{USA}
}

\author{Rocco Oliveto}
\affiliation{%
  \institution{University of Molise}
  \country{Italy}
}
\email{rocco.oliveto@unimol.it}

\author{Simone Scalabrino}
\affiliation{%
  \institution{University of Molise}
  \country{Italy}
}
\email{simone.scalabrino@unimol.it}

\author{Antonio Mastropaolo}
\email{amastropaolo@wm.edu}
\affiliation{%
  \institution{William \& Mary}
  \city{Williamsburg}
  \state{VA}
  \country{USA}
}

\renewcommand{\shortauthors}{Vitale \etal}

\newboolean{showcomments}
\setboolean{showcomments}{true}

\ifthenelse{\boolean{showcomments}}

\newmdenv[
  linecolor=black,
  linewidth=1pt,
  roundcorner=4pt,
  backgroundcolor=gray!10,
  innerleftmargin=10pt,
  innerrightmargin=10pt,
  innertopmargin=4pt,
  innerbottommargin=8pt,
  skipabove=6pt,
  skipbelow=6pt,
  shadowsize=3pt,
  shadowcolor=gray!20,
  middlelinecolor=black,
  leftmargin=0pt,
  rightmargin=0pt,
  align=center
]{iconbox}

\newcommand{\boxwithsideicon}[3]{%
  \begin{iconbox}
    \begin{minipage}[t]{\linewidth}
      \textbf{#2}%
      \hspace{0.5em}%
      \raisebox{-0.4\height}{\includegraphics[width=0.75cm]{#1}} \\[0.5ex]
      #3
    \end{minipage}%
  \end{iconbox}
}

\newtcolorbox{simpleacademicbox}{
  colback=gray!10,
  colframe=black!30,
  boxrule=0.3pt,
  arc=1pt,
  left=6pt,
  right=6pt,
  top=4pt,
  bottom=4pt,
}

%% Abstract
\begin{abstract}
Recent progress in Large Language Models (LLMs) has substantially advanced the automation of software engineering (SE) tasks, enabling complex activities such as code generation and code summarization. However, the black-box nature of LLMs remains a major barrier to their adoption in high-stakes and safety-critical domains, where explainability and transparency are vital for trust, accountability, and effective human supervision. Despite increasing interest in explainable AI for software engineering, existing methods lack domain-specific explanations aligned with how practitioners reason about SE artifacts.
To address this gap, we introduce \approach, the first fully automated, model-agnostic explainability framework tailored to software engineering tasks. Based on Shapley values, \approach attributes model outputs to high-level input features through systematic input perturbation and task-specific similarity comparisons, while remaining compatible with both open-source and proprietary LLMs. We evaluate \approach on two bi-modal SE tasks: code generation and code summarization.
The results show that \approach assigns less importance to irrelevant input features and produces explanations with higher fidelity than baseline methods. A practitioner survey involving \nparticipants participants shows that \approach helps practitioners better interpret model outputs and make more informed decisions. Collectively, \approach represents a meaningful step toward practical explainable AI in software engineering. \approach is available at \url{https://github.com/deviserlab/FeatureSHAP}.
\end{abstract}

%% CCS Concepts
\begin{CCSXML}
<ccs2012>
   <concept>
       <concept_id>10011007.10011074</concept_id>
       <concept_desc>Software and its engineering~Software creation and management</concept_desc>
       <concept_significance>500</concept_significance>
       </concept>
 </ccs2012>
\end{CCSXML}

\ccsdesc[500]{Software and its engineering~Software creation and management}

\keywords{Large Language Models for Code, Code Generation, Code Quality}

\maketitle

\section{Introduction}
\label{sec:intro}
Automation in software engineering (SE) is rapidly evolving due to two main drivers: the unprecedented volume and sheer amount of code data available from open-source ecosystems such as GitHub and the rise of Large Language Models (LLMs) capable of understanding and generating code \cite{chen2021evaluating, roziere2023code, lozhkov2024starcoder, huang2024opencoder, guo2024deepseek, hui2024qwen2}.
Among these, Neural Code Models (NCMs) represent a focused branch of LLM research--models specifically optimized for coding-related tasks, offering finer control and domain awareness in SE applications \cite{feng2020codebert, guo2020graphcodebert, wang2021codet5, guo2022unixcoder, shi2025sotana}.
Trained in a vast corpus of text enriched with large-scale open-source code, these models have achieved impressive performance in a wide range of software engineering applications, including code completion \cite{ciniselli2021empirical, chen2021evaluating}, program repair \cite{jin2023inferfix,fan2023automated,jiang2023impact,xia2023automated}, code summarization \cite{ahmed2022few, ahmed2024automatic, sun2024source, vitale2025optimizing}, and test case generation \cite{schafer2023empirical, lemieux2023codamosa, alagarsamy2025enhancing}.
Through scaling to billions and even trillions of parameters\footnote{For example, recent large-scale models such as OpenAI's GPT-4, Anthropic's Claude 3, and Google's Gemini 1.5 are reported to contain on the order of one to several trillion parameters, depending on the configuration disclosed.}, LLMs are capable of capturing complex relationships between code and natural language, allowing them to generalize effectively across programming languages, coding styles, and diverse development contexts.
This capability has placed LLMs at the core of commercial software engineering tools, such as GitHub Copilot \cite{githubcopilot}, which promise to enhance developer productivity throughout the software development life-cycle \cite{mo2025assessing, obrienprompt, terragni2025future}. This signals a fundamental shift in SE automation, where AI-driven models are no longer confined to experimental settings, but are actively embedded in IDEs and integrated into development workflows.

\begin{figure*}[t!]
  \centering
  \includegraphics[width=\linewidth]{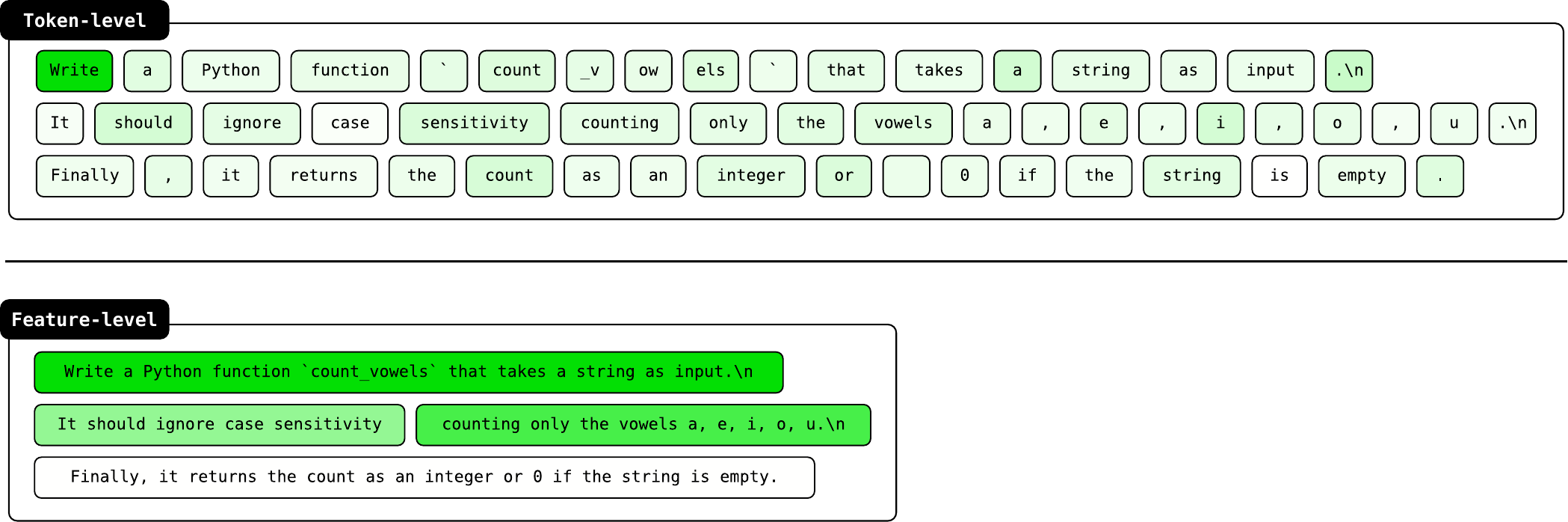}
    \caption{Token‐level (top) and feature‐level (bottom) attributions for the same input; darker rectangles represent higher attribution scores.}
  \label{fig:example_intro}
\end{figure*}

However, LLMs operate as black boxes. Given a suggestion (\eg a generated code block), a developer can either blindly trust the model, which actually results in a productivity improvement, or not, which implies double-checking the output and, thus, spending more time on the task. In addition, LLMs are non-deterministic, \ie the same input can result in different outputs. Such challenges pose significant barriers to the trustworthy adoption of such models. Ideally, developers should neither blindly trust the model -- because of non-determinism -- nor completely mistrust it -- since it would dramatically reduce the benefits of using LLMs in the first place. Instead, we conjecture that the best option would be an \textit{evidence-based trust}: predictions should be accompanied by easy-to-check pieces of evidence that report to what extent the user can trust the output and its parts.
Explainable AI (xAI) for SE methods emerge as a practical and promising solution to this problem. xAI methods reveal which inputs most influence an LLM's output, enabling practitioners to better assess the reliability of generated code, debug unexpected behavior, and refine prompts to align the model with the user's intended goals.
Among existing xAI techniques, two of the most popular are LIME \cite{ribeiro2016should} and SHAP \cite{lundberg2017unified}.
LIME \cite{ribeiro2016should} explains a prediction by generating perturbed versions of the input (\eg by masking or removing tokens) and training a simple interpretable model to approximate how these perturbations affect the generated output.
SHAP \cite{lundberg2017unified}, instead, estimates the importance of each input token by measuring how the inclusion or exclusion of that element changes the output across all possible subsets of tokens.
Existing explainability techniques typically operate at the level of individual tokens, which limits their applicability in software engineering contexts.

For example, consider the attributions for the prompt shown in \figref{fig:example_intro}. In the token-level view, the attribution is spread across many tokens, resulting in an explanation that is difficult to interpret: Developers do not reason one token at a time, and the distribution of importance across subwords offers little insight into which conceptual parts of the instruction actually contributed to model's output. Instead, the feature-level view provides a different perspective. Grouping tokens into semantically coherent units, such as the task description, the constraint on case sensitivity, and the definition of the vowel set, produces an explanation that aligns more closely with how developers understand and reason about prompts. This higher-level representation makes the explanation substantially more actionable.
For instance, when a specific feature receives negligible attribution (\eg the final requirement in the example), the developer may immediately infer two possibilities: Either the model ignored an intended constraint, which may imply a potential issue in the generated code, or the model produced the correct result regardless, explaining that the overlooked requirement was not contributing in the first place (\ie the requirement could have been omitted without affecting the model's behavior).
In both cases, feature-level attribution provides a clearer, more practical basis for explaining the model's behavior than token-level analysis and naturally points toward the need for xAI techniques that operate over semantically meaningful, task-dependent \textit{features}—units. When the input consists of source code, such units may align with program structures, such as method signatures, code blocks, documentation fragments, that define the semantics developers rely on when reasoning about the way the model processed and responded to the input.

To this end, we introduce \approach, a model- and task-agnostic explainability framework that estimates the importance of an input $i$ in generating the related output $M(i)$. Grounded in the theoretical foundations of Shapley values \cite{shapley1953value}, \approach provides a principled mechanism for estimating feature attribution beyond specific model architectures. Unlike recent efforts such as doCode \cite{palacio2024toward}, ASTrust \cite{palacio2024towards}, and CodeQ \cite{palacio2025explaining}, which focus on interpretability, \approach advances explainability by attributing model outputs to semantically meaningful, feature-level code constructs rather than token-level elements, producing explanations that better align with how developers reason about software artifacts.

\approach follows a three-step workflow. First, it decomposes the input into semantically meaningful units, termed \textit{features}, which capture higher-level abstractions. Second, it performs a series of controlled perturbation-based simulations, systematically including or excluding each feature or combination of features. Third, by observing the corresponding changes in the output of the model, \approach quantifies the contribution of each feature, producing attributions (a real value score in the range $[0, 1]$) that explain how and to what extent the output depends on different parts of the input. Higher attributions assigned to features indicate a greater influence on the output.

To capture the breadth of the output produced by LLMs, we evaluated \approach on two representative tasks: code generation and code summarization. Together, they span the spectrum from the production of executable code to the generation of technical natural language, illustrating the generality of our approach.

In particular, we first try to understand whether \approach can detect parts of the prompt that surely do not influence the output. To this end, given a model $M$, we consider a set of prompts $I$, infer the results $O = \{M(i)~\forall~i \in I\}$, and then introduce extraneous features to the results $I_n = \{\text{inject}(i,ext(i))~\forall~i \in I\}$. Such features ($ext(i)$) are chosen so that they are surely non-influential, \ie by making sure that the output of the model does not change ($O_n = \{M(i)~\forall~i \in I_n\} = O$. By construction, a realiable xAI approach should attribute a very low score to such extraneous features.
Our results show that \approach consistently assigns negligible attribution scores to injected irrelevant features across all models and tasks, outperforming both random and LLM-based attribution baselines in distinguishing non-influential inputs.

In the second investigation, we examine whether \approach can provide human-centered and actionable support to developers by providing transparent explanations. To this end, we conducted a survey with \nparticipants professionals who regularly employ LLMs in their daily programming activities. Participants were asked to assess \approach explanations in code generation and summarization tasks, evaluating their faithfulness and usefulness to support practical use. Our results indicate that practitioners considered the attributions produced by \approach meaningful, enhancing their confidence and decision-making when assessing model generated outputs.
The combined findings of such two studies confirm both the technical reliability and the practical utility of \approach for explainable AI in software engineering.

To summarize, this work advances the state-of-the-art in explainable AI methods for SE via the following key contributions:
\begin{itemize}
\item It proposes the first fully automated, model-agnostic, and black-box explainability framework explicitly designed for software engineering tasks.
\item It provides empirical evidence supporting the usefulness of \approach in (i) finding irrelevant features in the prompt, and (ii) helping developers take decisions on the received recommendations.
\end{itemize}

In the remainder of this paper, we begin by providing the necessary background to help readers better understand the details of our proposed technique. We then introduce \approach, detailing the key components specifically designed to enable concept-level (feature-wise) explainability.
\secref{sec:design} describes the experimental design, while \secref{sec:results} presents and analyzes our findings through both quantitative metrics and a qualitative user study.
In \secref{sec:implications}, we situate our work within the broader landscape of explainable and interpretable software engineering, and reflect on the new opportunities unlocked by \approach.
We conclude by discussing the threats to validity associated with our study and summarizing the main takeaways.

All code and data used in our study are publicly available \cite{replication}, together with the associated tool at \url{https://github.com/deviserlab/FeatureSHAP}.

\section{Background \& Related Work}
\label{sec:motivation}
We first discuss the background of the concepts of explainability and interpretability. Then, we present related works specifically in the context of software engineering.

\subsection{Explainability and Interpretability}
There are two families of methodologies that can be adopted to make LLMs less opaque: interpretable methods and explainable methods \cite{lyu2024towards, zhao2024explainability, mersha2024explainable}.
Interpretable methods are those where the model or its features can be directly inspected by humans.
A classic example is linear regression, where model coefficients explicitly quantify the contribution of each input feature to the prediction.
By extension, in the SE domain, an interpretable method might involve the examination of specific parts or features of the models. For example, Sharma \etal \cite{sharma2022exploratory} examined the Transformer's \cite{vaswani2017attention} \footnote{Transformers are a deep learning architecture based on self-attention mechanisms, which power modern LLMs due to their ability to capture long-range dependencies and represent complex relationships in sequential data such as natural language and source code.} attention weights in a code-pretrained BERT \cite{devlin2019bert} model to identify which categories of tokens (\eg identifiers, separators) receive the most attention, suggesting their importance for the model's predictions.
In essence, an interpretable method aids in understanding the input–output relationship when certain conditions are met: (i) the model is transparent, meaning its internal structure and parameters can be directly examined; and (ii) the model or its features exhibit traceable behavior, such that specific patterns in the input can be logically linked to corresponding outputs through an understandable decision process.

Explainable methods refer to \textit{post-hoc} techniques that take a complex black-box model and produce an explanation for its outputs. 
Notably, they place no requirement for the model to be open-source or internally accessible-making them particularly valuable when working with proprietary or API-based systems. Common general techniques include LIME \cite{ribeiro2016should} and SHAP \cite{lundberg2017unified} that follow this paradigm by perturbing input features and observing corresponding changes in output to estimate the importance of features.

Given the focus of our paper on \textit{explainability} rather than \textit{interpretability}, we will report on the post-hoc explainability techniques that have been proposed in the literature.

LIME (Local Interpretable Model-agnostic Explanations) \cite{ribeiro2016should}, focuses on explaining why a model provides a particular output given a single input leveraging an interpretable model that learns to mimic the model's behavior when exposed to small perturbations of that input.
It works by creating a set of synthetic inputs that are slight variations of the original one. For example, in the context of text generation, it randomly removes or alters some tokens. Then, these perturbed inputs are given to the model to quantify how the generation changes. Let us consider the case in which a prompt asks a model to generate code to sort a list: If removing the word \texttt{\{sort\}} from the prompt causes the model to generate something completely different, LIME will infer that \texttt{\{sort\}} was important to generate the original output. Repeating this procedure through many perturbed examples and their corresponding predictions, LIME fits a simple, interpretable model (such as a linear model) with these data points. The weights learned by the surrogate model explain which parts of the original input were most important in steering towards the original generation.

SHAP (SHapley Additive exPlanations) \cite{lundberg2017unified} is based on cooperative game theory. SHAP assigns each input feature an importance value based on how much it contributes, on average, to the prediction, measured considering all possible combinations in which tokens could be added to the model input.
Intuitively, the model can be imagined as a team project, and each input feature can be imagined as a team member. SHAP evaluates how the final outcome would change as each team member is part of the group, averaging all possible orders in which they could be included. For example, if a code prompt contains both \texttt{\{sort\}} and \texttt{\{reverse\}}, SHAP asks: ``What value does adding \texttt{\{sort\}} provide to the model's output, when considered alongside every possible subset of the other features?'' By systematically averaging these contributions, SHAP assigns each feature an overall importance score. SHAP is the only method that provides additive feature attributions while guaranteeing local accuracy, missingness, and consistency properties.
However, calculating the exact SHAP values requires evaluating the model on every possible subset of input tokens, which becomes infeasible as the input size grows.

Recently, to meet LLMs in the explainability context, TokenSHAP \cite{horovicz2024tokenshap} has been proposed. Like SHAP, TokenSHAP is grounded in cooperative game theory, treating each token as a ``player'' and estimating how much each contributes to the model's output. The intuition is straightforward: By systematically removing tokens and observing how the model's response changes, TokenSHAP can estimate the average effect that each token has on the original generation.
Since evaluating every possible combination of tokens is computationally infeasible for real-world prompts, TokenSHAP employs a Monte Carlo sampling strategy to approximate the Shapley value for each token efficiently. For a given prompt, it generates many variants of the prompt by omitting different tokens, queries the LLM for each, and measures how similar each response is to the full prompt TF-IDF representations of the outputs. The average change in response similarity, when a specific token is included versus excluded, determines the token importance score. For example, if the word \texttt{\{sort\}} is removed from a code generation prompt and the model does not suggest (again) sorting operations, TokenSHAP will assign a high importance to \texttt{\{sort\}} token.

Building on token-level approaches, SyntaxSHAP \cite{amara2024syntaxshap} has been proposed to incorporate higher-level linguistic structure into explanations. Like SHAP, it is grounded in cooperative game theory, but instead of forming arbitrary subsets of tokens, SyntaxSHAP constrains coalitions to follow the dependency parse tree of the input text. This ensures that related tokens are treated as coherent units. For instance, given the prompt \texttt{\{sort the list in descending order\}}, SyntaxSHAP evaluates the phrase \texttt{\{descending order\}} as a single unit when estimating its contribution to the output, moving beyond token-based attributions and produces explanations that are more syntactically faithful and coherent. However, the reliance on surface syntax means that the explanations still diverge from human expectations when semantic concepts are overlooked, leaving open the broader challenge of aligning explanations with domain-specific reasoning needs.

\subsection{Explainability and Interpretability in Software Engineering}
In software engineering, efforts toward making neural models more transparent have grown alongside the adoption of large language models for code-related tasks. Early work on model transparency in SE primarily emphasized interpretability--that is, understanding how models make predictions--rather than explainability, which focuses on why specific predictions occur and how they relate to human reasoning processes \cite{1034239}.

Among recent contributions, Palacio \etal introduced several interpretability frameworks for neural code models. Their work on \textit{do$_{code}$} \cite{palacio2024toward} proposes a post-hoc interpretability approach based on causal inference, designed to reveal how specific code properties influence model decisions. Through case studies on diverse deep learning architectures, they show that \textit{do$_{code}$} can expose model sensitivities to syntactic variations and help identify potential confounding factors in code representations.

In the follow-up work, the same authors presented \textit{AST$_{rust}$} \cite{palacio2024towards}, which explains the model predictions by associating confidence values with syntactic structures in Abstract Syntax Trees (ASTs). Unlike token-based saliency methods, \textit{AST$_{rust}$} grounds explanations in language-specific constructs, bridging model behavior, and programming syntax.

Most recently, \textit{CodeQ} \cite{palacio2025explaining} was proposed as a rationale-based interpretability method that identifies minimal input subsets--termed code rationales--responsible for model predictions. Through exploratory analysis and a user study, \textit{CodeQ} was shown to produce informative, human-readable explanations and facilitate comparisons between model and developer reasoning.

Collectively, these works have advanced interpretability for neural code models, yet they share important limitations. Each approach requires access to internal model states or token-level confidence scores, restricting their applicability to black-box or proprietary models and making their performance dependent on model calibration. Furthermore, most existing techniques remain anchored at the token or sub-token level, yielding dense visualizations that highlight low-level text fragments rather than semantically meaningful code structures such as function bodies, documentation blocks, or API calls. As a result, their explanations often diverge from the way developers conceptualize and review software artifacts.

Finally, the metrics commonly used to quantify feature importance--such as output-probability shifts or surrogate-model weights--are largely domain-agnostic and overlook the semantics of software engineering tasks. 

Moreover, token-level attributions are inherently difficult to interpret and reason about, often resulting in dense and fragmented heatmaps that obscure rather than clarify model behavior \cite{cao2025systematic}. To address this limitation, \approach introduces an abstraction layer over fine-grained token attributions, aggregating them into higher-level features that correspond to semantically meaningful program elements or documentation components. This abstraction improves readability and bridges the gap between low-level model behavior and the conceptual structures that practitioners naturally employ when reasoning about software systems. Furthermore, \approach is a model-agnostic explainability technique that operates only on observable input–output behavior, allowing the generation of explanations even when model internals are inaccessible. In doing so, it broadens applicability to proprietary or black-box systems and strengthens trust in real-world production-level environments. Collectively, these characteristics highlight the need for explainability approaches such as \approach--methods that function at a higher semantic level, remain model-agnostic, and align with how developers naturally understand and interact with software.

\section{\approach}
\begin{figure*}[t!]
  \centering
  \includegraphics[width=\linewidth]{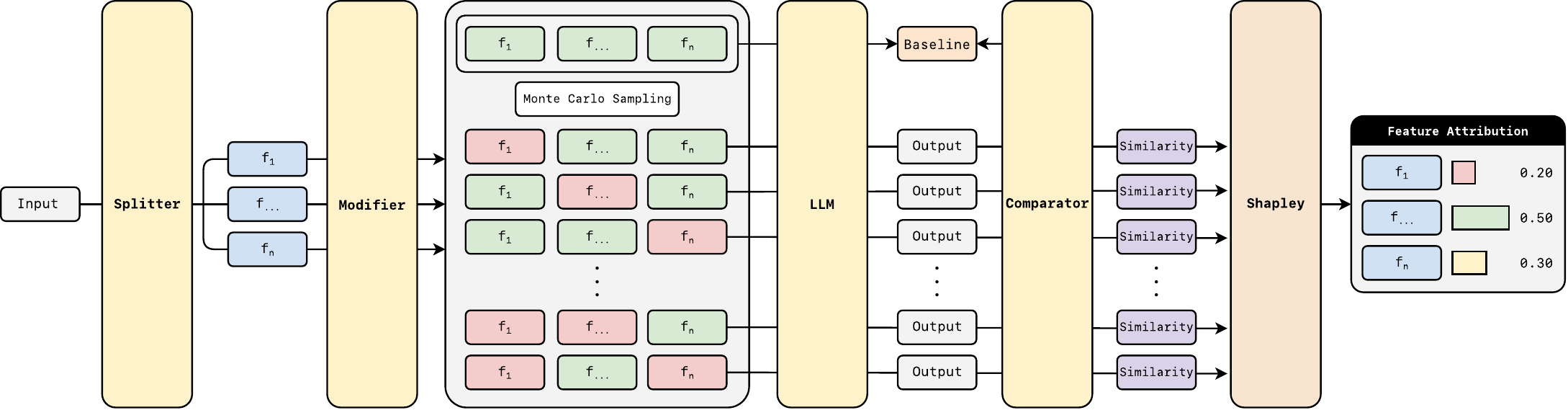}
  \caption{FeatureSHAP pipeline.}
  \label{fig:featureshap-pipeline}
\end{figure*}

To overcome the core limitations of state-of-the-art techniques, we propose \approach, a flexible, black-box approach designed specifically for software engineering tasks. Built on the foundations of TokenSHAP \cite{horovicz2024tokenshap}, \approach goes beyond simple token attributions by allowing the input to be partitioned into semantically meaningful features \ie documentation blocks or code sections, that better align with how developers naturally interpret code. This shift from treating each token as a "player" in the Shapley framework to grouping tokens into higher-level, domain-relevant features enables explanations that are both more interpretable and computationally efficient. As a result, \approach provides fine-grained task-adaptable attributions that better match the real-world reasoning of software engineering practitioners.

From a high-level perspective, \approach takes a given input $i$, a model $M$, and the output generated by the model given the input $o = M(I)$, and returns (i) the \textit{features}, \ie a partition of the tokens identified in $i$ ($F_i \in \mathcal{P}(\text{tokenize}(i))$, and (ii) an \textit{attribution function} for each of them $\alpha: f_i \rightarrow \Re$. The higher the attribution assigned to a given feature ($\phi(f \in F_i)$), the more important the feature is for the model $M$ to generate $o$ from $i$.

The \approach approach consists of three main components: \textit{splitter}, \textit{modifier}, and \textit{comparator}. Each of these can be adapted or instantiated to fit the demands of a particular task at hand. In the following, we first present the roles of the three components and then explain the process followed by \approach to compute attribution values for each feature.

\textbf{The Feature Attribution Process in \approach.}
 In \figref{fig:featureshap-pipeline} we report on the inner working of \approach.
Given an input $i = \langle t_1,~t_2,~\dots,~t_n \rangle$, where $t_j$ is the $j$-th token, the process begins by passing $i$ to the \textit{splitter} module. Such a module determines how the input is partitioned into features. Formally, the splitter is a function $\sigma$ that, given a sequence of tokens extracted from the input $i$ ($\text{tokenize}(i)$), returns a specific partition of this sequence, \ie $\langle f_1, \dots, f_n\rangle$, where each $f_i$ is a subsequence of tokens in $i$. The splitter could theoretically be as simple as a line or sentence splitter, but ideally it should identify a partition of tokens (features) so that each feature is \textit{semantically} meaningful to a developer. 
Once features are identified, the \textit{modifier} module implements how the input is altered to assess the impact of some specific feature, by mirroring the canonical removal step in the Shapley value framework. Modifier modules are based on an alteration function $\mu$ that, given a feature $f_i$ resulting from the use of the splitter module, returns an altered version of such a sequence, $f'_i$. The simplest alteration function is one that always removes the feature from the input \ie $\mu_r(f) = \langle\rangle$. Ideally, the modifier module should alter all the possible combinations of features ($\mathcal{P}(\sigma(i))$). Since this is very expensive in practice, to efficiently cover the space of all possible feature subsets, we use Monte Carlo sampling \cite{metropolis1949monte} as in TokenSHAP \cite{horovicz2024tokenshap}.
For each chosen set of features to alter, the modifier module alters them and leaves the others as in the original input. For example, given the input $i$ and its features $\sigma(i) = \langle f_1, f_2, f_3, f_4\rangle$, if the sample includes two sets of features to be altered, \ie $\{\{f_2\},~\{f_3, f_4\}\}$, the modifier will produce two altered versions of $i$, \ie $i'_1 = \langle f_1, \mu(f_2), f_3, f_4 \rangle$ and $i'_2 = \langle f_1, f_2, \mu(f_3), \mu(f_4) \rangle$.
Finally, the \textit{comparator} module defines the value function for the Shapley game, \ie measures how the model's output changes when the altered inputs $i'_j$ are used instead of the original ones $i$. Formally, the comparator is a similarity function $s$ that, given the output before the alteration ($M(i)$) and the one obtained with one of the altered versions $i'_j$ ($M(i'_j)$), measures to what extent they are close. Two completely different outputs will have score 0, while identical ones will have score 1. An example of a comparator function is the classic TF-IDF cosine similarity used in previous work \cite{horovicz2024tokenshap}.
All computed similarity scores are collected and fed to the \textit{Shapley engine} that estimates the contribution of each feature $f_i$. By analyzing how similarity changes when features are included or omitted in different combinations, the Shapley engine produces a set of attribution scores $(\phi_1,\ldots, \phi_m)$ for each feature.

\textbf{Instantiation of \approach.}
In this work, we instantiate \approach by making specific design decision in terms of the \textit{splitter}, \textit{modifier} (specifically, the altering function), and \textit{comparator} components for the tasks we aim to tackle in the experiment, as we will explain later, \ie code generation and code summarization.
First, since the \textit{splitter} strictly depends on the input type and even task at hand, we defined two different splitters: one for tasks that require natural language text as input (\eg code generation) and one for tasks that require source code as input (\eg code summarization).

\begin{figure}
\footnotesize
\begin{tcolorbox}[title=\textbf{Prompt}, colback=gray!5, colframe=black,
width=0.9\columnwidth,boxrule=0.15mm]
\texttt{SYSTEM\_PROMPT:} {You are a docstring segmenter. Split the user given docstring into granular sections.} \\

\texttt{CRITICAL REQUIREMENTS:}
\begin{itemize}
    \item {You MUST always output a list of strings - no other format is acceptable}
    \item {You MUST NOT paraphrase, rewrite, or modify any text - only extract exactly as written}
    \item {Each string must be an exact substring from the original docstring}
    \item {Include all original whitespace, newlines, and formatting exactly as they appear} \\
\end{itemize}

\texttt{JSON FORMATTING REQUIREMENTS:}
\begin{itemize}
    \item Output ONLY valid JSON - an object with numeric keys and string values:  \texttt{\{``0'': ``string1'', ``1'': ``string2'', ``2'': ``string3''\}}
    \item {Each string value must be a meaningful coding sentence segments.}
    \item Every key must be a string representing the index: ``0'', \texttt{``1'', ``2'', etc.}
    \item {Every value must be a string segment enclosed in double quotes}
    \item {Ensure all quotes and special characters are properly escaped} \\
\end{itemize}

{Here are examples, your response should follow this format:}\\
\texttt{\{ICL Example 1\}}\\
\texttt{\{ICL Example 2\}}\\
\texttt{\{ICL Example 3\}}\\

\texttt{USER\_INPUT:} Docstring: \texttt{\{docstring\}}
\end{tcolorbox}
\caption{Prompt for the LLM-based splitter (code generation). Here, \texttt{\{docstring\}} refers to the docstring-formatted prompt used by BigCodeBench for code generation.}
\label{fig:prompt_splitter}
\end{figure}

\begin{figure}
\footnotesize
\begin{tcolorbox}[title=\textbf{ICL Example 1}, colback=gray!5, colframe=black,
width=0.9\columnwidth,boxrule=0.15mm]
\texttt{INPUT:} \\
"Search for a specific string within the JSON data of files in a given directory and its subdirectories. This function recursively scans the specified directory for JSON files, then checks each file to see if the given string is present within the JSON data structure.\\Note that: The string search is case-sensitive and looks for a match within the structure of the JSON data, not just as a substring in the file content. If the directory does not contain any JSON files or if no JSON files contain the string, an empty list is returned.\\The function should output with:\\    list: A list of file paths (str) containing the string within their JSON data." \\

\texttt{OUTPUT:} \\
\{"0": "Search for a specific string within the JSON data of files in a given directory and its subdirectories. ",\\ "1": "This function recursively scans the specified directory for JSON files, ",\\ "2": "then checks each file to see if the given string is present within the JSON data structure.",\\ "3": "Note that: ",\\ "4": "The string search is case-sensitive and looks for a match within the structure of the JSON data,",\\ "5": " not just as a substring in the file content. ",\\ "6": "If the directory does not contain any JSON files or if no JSON files contain the string, an empty list is returned.",\\ "7": "The function should output with:",\\ "8": "    list: A list of file paths (str) containing the string within their JSON data."\}
\end{tcolorbox}
\caption{Demonstration of an ICL example used in LLM-based splitter.}
\label{fig:icl_example}
\end{figure}

As for the former $\sigma_n$, we use a few-shot in-context learning approach where an LLM splitter segments input prompts into high-level features such as summary, detailed description, and the function signature, based on human-annotated examples. We use \texttt{gpt-4.1-mini} \cite{gpt41} as the engine behind the LLM splitter to balance both performance and cost. We use a prompt engineered to ensure the high quality outputs described in Figure \ref{fig:prompt_splitter}.
In detail, we enforce a rigid output format: the LLM must return a JSON object where each key is a string index ("0", "1", …) and each value is a feature representing an exact substring of the original docstring, including all whitespace and formatting. We constrain the model to produce outputs that are lossless with respect to the input, thus preventing paraphrasing and ensuring that the segmented units can be leveraged as structured features. We do this by verifying that the ordered concatenation of all extracted segments reproduces the original docstring exactly.
A demonstration of the ICL examples used by the splitter can be observed in Figure \ref{fig:icl_example}.

As for the splitter that treats the source code $\sigma_c$, we leverage the Tree-Sitter parser\footnote{\url{https://tree-sitter.github.io/tree-sitter/}}.
First, we build the abstract syntax tree (AST) of the method, denoted $\mathcal{T}(i)$.
Then, since the parser operates on the UTF-8 encoded form of the snippet, we use byte offsets to identify the relevant code blocks.
From $\mathcal{T}(i)$, we extract the method declaration ($s_0$) together with the ordered list of its top-level statements $(s_1, s_2, \dots, s_k)$.
Each of these nodes contains a starting byte position $p(s_i)$ in the encoded input $i$.
We collect all such positions, including the end-of-input position $p(s_{k+1})$, sort them in ascending order, and produce the $i$-th block by slicing the encoded text from $p(i_j)$ to $p(s_{i+1})$.
Formally, $\sigma_c(i)$ returns the features $\langle f_0,\dots,f_k\rangle$ where $f_i=\operatorname{slice} \bigl(i,\,p(s_i),\,p(s_{i+1})\bigr)$ for $i=0,\dots,k$.

As for the \textit{modifier}, we decided to retain the simplicity and rigor of SHAP's removal operation. Therefore, we adopt the previously-mentioned altering function $\mu_r(f_i) = \langle\rangle$ which removes tokens related to a given feature $f_i$.

Finally, the \textit{comparator} depends on the type of artifact produced by the target model $M$ in the task, which is again source code or natural language text.
As for the former, we use CodeBLEU \cite{ren2020codebleu}, a metric designed to assess the similarity of candidate and reference code considering (i) the n-grams overlap through a weighted BLEU \cite{papineni2002bleu}, (ii) the syntax through Abstract Syntax Tree, and (iii) the semantics via data-flow. 
For the latter, instead, since the output is natural language text, we adopt BERTScore \cite{zhang2019bertscore}, an effective metric to measure semantic similarity between textual contents through contextual embeddings.

\begin{figure}
  \centering
  \includegraphics[width=\linewidth]{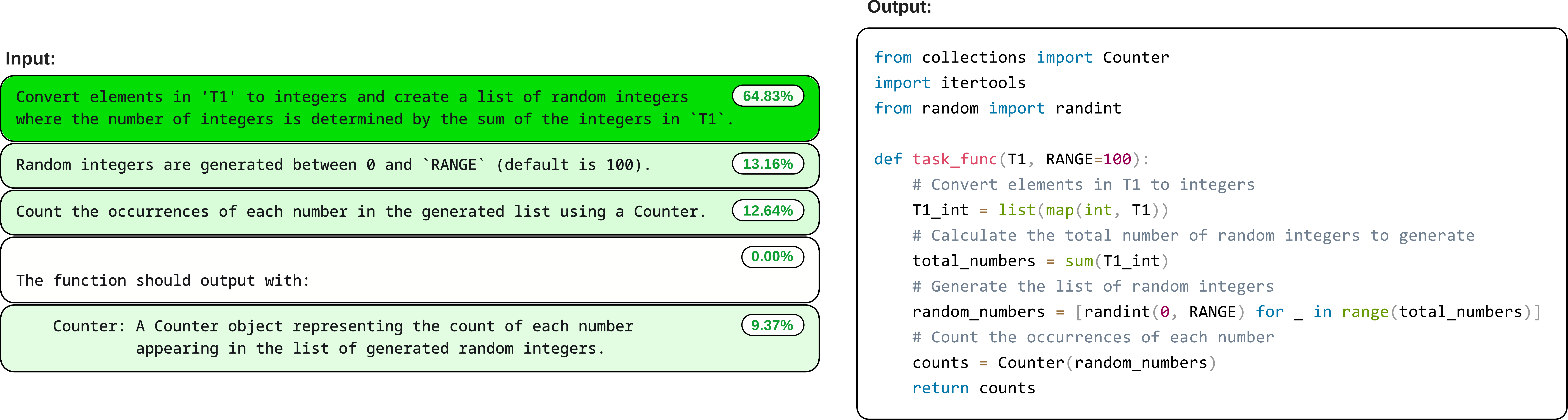}
  \caption{\approach feature attribution for code generation. The darker the highlight, the higher the influence of that feature on the generated code.}
  \label{fig:code-gen-feature}
\end{figure}

\textbf{\approach in Action.} We first provide a concrete example to help the reader intuitively grasp the capabilities of \approach. \figref{fig:code-gen-feature} illustrates a feature attribution scenario for a code generation task. The left portion shows the input prompt, while the right one displays the code generated by the LLM. \approach evaluates the contribution of each semantically meaningful feature in the input and visualizes their relative influence using varying shades of green -- darker shades indicate a higher contribution to the generated output. In this example, the top two features alone account for over $\sim$78\% of the total attribution budget (normalized to one), indicating that a significant portion of the model's behavior can be explained by a small set of semantically rich input segments highlighting the core behavior of the underlying method. \\

\textbf{\emph{Computational Efficiency}}. The computational cost of \approach depends mainly on two factors: (i) the number of features produced by the splitter, and (ii) the Monte Carlo sampling ratio used to approximate Shapley values, which determines the percentage of the total number of combinations of features that will be sampled. Because features are larger semantic units, \approach usually requires far fewer evaluations than token-level methods such as TokenSHAP or LIME. For example, an input of 200 tokens might be grouped into only 8 features, shrinking the theoretical space from $2^{200}$ to $2^{8}$. Monte Carlo sampling might then further reduce the number of coalitions (\ie subsets of features) that must be evaluated when the sampling ratio is lower than 1, making the approach tractable in practice. On the other hand, the comparators (\eg CodeBLEU or BERTScore) introduce extra overhead, since each perturbed output must be scored with a non-trivial metric. In summary, \approach theoretically strikes a better balance between explainability and efficiency than token-level attribution, but its runtime remains non-trivial and sensitive to the complexity of some of the design decisions made in terms of splitter, modifier, and (especially, in our specific instantiation) comparator.

\section{Study Design}
\label{sec:design}
The \textit{goal} of our study is to evaluate whether \approach is both technically effective--accurately determining attribution values for the identified features--and practically acceptable to developers who rely on such explanations. These two dimensions are closely interrelated, since an explainability method must be both reliable and usable to achieve real impact. Consequently, we define the following research questions.

\begin{itemize}
 
 \item[\bf \RQ{1}:] \textit{To what extent does \approach assign low attribution to randomly injected non-contributing features?}
  With this research question we aim to assess to what extent \approach can reliably assign low attribution to random noise features that are irrelevant to the model's output.

  \item[\bf \RQ{2}:] \textit{To what extent does \approach assign low attribution to realistic non-contributing features?}
  With this research question we aim to assess to what extent \approach can assign negligible attribution to realistic injected noise that resemble valid features but are designed to be irrelevant.

  \item[\bf \RQ{3}:] \textit{To what extent does \approach facilitate the decision-making process in real-world settings?}
  
  This research question aims to evaluate whether the attributions produced by \approach are both faithful to the model's decision process and useful to practitioners.
  \end{itemize}

\subsection{Context Selection}
\label{sec:context}
The context of our study is composed of \textit{objects} and \textit{subjects}.
The \textit{objects} are LLMs and benchmarks, necessary to answer \RQ{1}, while the \textit{subjects} are human participants in a survey that we conduct to answer \RQ{2}. In the following, we detail the choice of all such elements.

\textbf{LLMs.}
\label{sec:models}
As explained, \approach is model-agnostic. Thus, we want to demonstrate its effectiveness on two classes of models: \textit{open-weights} and \textit{closed-weights}.
We select LLMs that are representative of the state-of-the-art and widely used in software engineering automation practices. In particular, we focus on \texttt{Qwen2.5-Coder} \cite{hui2024qwen2} and \texttt{GPT-4.1-nano} \cite{gpt41}.
The former is a family of code specialized LLMs with state-of-the-art performance on code benchmarks, and it is our representative open-weight model series. We experiment with the three different sizes (3B, 7B, and 32B) to study the effects of the size of the model on the \approach attribution score.
The latter is a lightweight variant of the \texttt{GPT-4.1} family. We choose \texttt{GPT-4.1-Nano} as it balances the strong performance of GPT-based models on code-specific tasks with a manageable cost and runtime for practical evaluation. For the implementation, we conduct the validation previously described using the models presented in \secref{sec:context}. Inference is performed with greedy decoding and a temperature of 0 to foster reproducibility. For the \texttt{Qwen2.5-Coder} models, we leverage vLLM \cite{kwon2023efficient} to enable efficient inference. We apply Monte Carlo sampling within \approach using both \texttt{Qwen2.5-Coder} and \texttt{GPT-4.1-Nano}, setting the sampling ratio to~0 to balance computational cost.

\textbf{Tasks and Benchmarks. \label{sec:benchmark_selection}}
We consider two representative and complementary software engineering tasks. The first is \textit{code generation}, which involves synthesizing executable source code from natural language specifications. The second is \textit{code summarization}, which focuses on producing concise natural language descriptions from source code. Together, these tasks capture the two fundamental directions of code–text interaction: generating code from text and generating text from code. To ensure experimental diversity and alignment with established benchmarks in the literature, we rely on input prompts from \texttt{BigCodeBench} \cite{zhuo2024bigcodebench} to evaluate code generation and from \texttt{CodeSearchNet} \cite{lu2021codexglue} to evaluate code summarization.
We select such benchmarks, as they are representative of their task and have become widely adopted in recent literature \cite{hui2024qwen2,ahmad2025opencodeinstruct,wu2025teaching,wang2025towards,huang2025opencoder} \cite{cao2025rethinking,fang2025enhanced,sontakke2022code,geng2023interpretation}.\\

\noindent \texttt{BigCodeBench} \cite{zhuo2024bigcodebench} is a large-scale Python code generation benchmark designed to assess LLMs in realistic programming scenarios widely used in recent research \cite{hui2024qwen2, abouelenin2025phi, lambert2024tulu}. It includes 1,140 individual, function-level tasks that span 7 diverse domains and invoke APIs from 139 widely-used Python libraries.
We use the \texttt{BigCodeBench-Instruct} set in our experiments using the Huggingface's \texttt{datasets} package. This benchmark enables us to assess \approach on function-level tasks typically encountered in real-world software development. \\

\noindent \texttt{CodeSearchNet} \cite{Husain2019CodeSearchNet} is a code summarization benchmark modified and prepared by Makharev \etal \cite{makharev2025code}. We choose this dataset since (i) it is part of \texttt{CodeXGLUE} \cite{lu2021codexglue}, an established widely studied benchmark in the literature \cite{ahmed2022few,shi2022we,afrin2025resource,virk2025calibration} and (ii) this version of \texttt{CodeSearchNet} further cleaned and processed by Makharev \etal \cite{makharev2025code}, consisting of 846 python \texttt{<code,summary>} pairs. This ensures that we can evaluate \approach on a high quality dataset with a reasonable number of samples.

Note that in this context, we do not need to run test cases (\eg for code generation) or check the similarity to ground truth (\eg for code summarization, since the aim of explaining a model's output is independent of the correctness of the generation. However, we still select state-of-the-art benchmarks, as they contain clean and reliable input \cite{zhuo2024bigcodebench,makharev2025code}.\\

\textbf{Participants.}
We recruited \nparticipants participants from the academic network of the authors (\textit{convenience sampling}). We only included software engineers, doctoral students, postdoctoral researchers, and early-career faculty -- all of them with solid backgrounds in software engineering or machine learning. Our main selection criteria were two: (i) the participant should be familiar with LLM-based development tools and (ii) they should be capable of critically assessing model explanations. Such criteria ensure that the feedback collected was both informed and reliable.

\subsection{Experimental Procedure for \RQ{1}}
To answer \RQ{1}, we develop a noise injection methodology based on Horovicz \etal \cite{horovicz2024tokenshap}. The procedure consists of three steps. First, we perform \textit{noise injection} into original prompts from the benchmarks described in Section \ref{sec:benchmark_selection}. Second, we \textit{filter the samples} and retain only the cases where model output remains unchanged after noise injection, ensuring that the injected tokens are genuinely irrelevant to the generation. Third, we quantify the attribution gap by computing a \textit{Noise Score}, defined as the attribution assigned to injected noise features across the filtered instances.
If \approach works as intended, it should consistently assign an attribution score close to 0 to the irrelevant injected noise.
Finally, following Horovicz \etal ~\cite{horovicz2024tokenshap}, we compare \approach against a \textit{random-based attribution} baseline.

\textbf{Noise Injection.}
 Consider a prompt $i$ that is being tested among the benchmarks. First, we identify the features of $i$ using the compatible splitter of \approach based on the type of input ($\sigma_n$ for natural language and $\sigma_c$ for source code). We obtain a sequence of features, $\langle f_1, \dots, f_n \rangle$. Then, we select a noisy feature $nf$ using the strategy we will describe later and inject it at a random position. To ensure that injected noise is comparable to the features within the prompts, we make sure that $nf$ is about the same length as the other features, allowing at most a three-total difference.

In detail, the strategy we adopt in this setting is the \textit{nonsensical} injection.
We insert an out-of-distribution sentence selected from a pool of 20 seed sentences generated using GPT.
To build this pool, we prompted GPT to produce self-contained statements that are syntactically well-formed but conceptually incoherent.
We manually reviewed the pool to ensure that none of them expresses an actionable instruction, a programming task, or any form of meaningful guidance that could plausibly influence the model's behavior.
The outcome is a set of linguistically plausible sentences that are irrelevant to the task and therefore serve as genuine noise when injected into a prompt.
This setting evaluates whether \approach correctly assigns negligible attribution to content that cannot influence the behavior of the model by construction.
An example of such noise is the feature ``\textit{Zebras navigate labyrinthine caves using sonar echoes that humans cannot perceive}'' inserted into a sorting function prompt.

\textbf{Instances Filtering.} In noise-injection, to ensure the noise feature is truly irrelevant, we must make sure it does not contribute to the model's output by construction. Therefore, we only conduct the evaluation on cases where noise features are genuinely irrelevant to model output.
Our output filtering process retains only instances where injected noise does not alter the behavior of the model. In particular, for each instance of each benchmark, we perform the following steps. First, we generate the baseline output $o = M(i)$ from the original prompt $i$. Second, we create the noisy version of the prompt (\textit{nonsensical} injection), obtaining the variants $ni_{\text{rand}}$. We then run the model on the variant, obtaining the candidate output $O' = M(ni_{\text{rand}})$. Specifically, we retain the pair $\langle ni_{rand}, M(ni_{rand})\rangle$ only if $O' = o$, \ie the injected noise does not change the original output. To this aim, we use exact match to make sure that the injected noise has no impact whatsoever on the output (neither from a functional nor from a non-functional point of view). All other variants are discarded.
To have a tractable computation time for the calculation of exact attribution scores using \approach, we filter out instances with more than 12 features since computing the exact Shapley values is computationally expensive (it requires to consider all possible feature coalitions of $2^{|F|}$. With this step, we filtered approximately 13\% of the samples from \texttt{BigCodeBench} and 6\% from \texttt{CodeSearchNet}.

\textbf{Metrics.} 
We measure \approach's capability to assign low attribution scores to non-contributing features using the \textit{Noise Score} metric, which quantifies, for each instance, the attribution assigned to the injected noise feature.
Specifically, given an instance $I$ with an injected noise feature $nf$, we compute its \textit{Noise Score} as: $S_I = A(nf)$ where $A(nf)$ denotes the attribution score assigned to the noise feature $nf$.
A lower value of $S_I$ indicates that the method successfully minimizes the importance assigned to non-contributing features, while higher values suggest that it incorrectly attributes relevance to them.  

We compute the distribution of \textit{Noise Scores} for all evaluated instances $\mathcal{D}$ in terms of mean and median.  
To statistically compare \approach against the baselines (more on this later), we perform pairwise tests on the per-instance \textit{Noise Scores} using the Wilcoxon signed-rank test \cite{wilcoxon1992individual}, and Cliff's delta effect size \cite{grissom2005effect}. For multiple comparisons, $p$-values are adjusted following Holm's correction \cite{holm1979simple}.

Note that we focus our evaluation on noise features only. We do this since those are the only elements for which non-contribution can be established with certainty, given our construction and filtering criteria. In contrast, original features from the prompt may naturally receive low or even zero attribution without letting us be able to determine whether they truly contribute or not.

\textbf{Random Baseline.}
We compare \approach with a \textit{random-based attribution baseline}.
The latter assigns random attribution scores between 0 and 1 to each feature. Since the sum of attributions should be 1, we normalize such scores dividing them by the sum of the randomly-assigned scores. This serves as a lower-bound performance indicator that establishes the minimum threshold that any meaningful attribution method should exceed.

\subsection{Experimental Procedure for \RQ{2}}
To answer \RQ{2}, we develop a cross-sample injection methodology that evaluates \approach under more realistic and challenging conditions. While \RQ{1} serves as a foundational check, verifying that \approach correctly assign low attribution to irrelevant out-of-distribution injected noise features, \RQ{2} assess whether \approach can do the same in a more realistic and challenging setting.
As for the previous, we first inject a noise feature, then filter the samples that allow model outputs to remain unchanged after noise injection, and finally compute the Noise Score.

\textbf{Noise Injection.}
For this \RQ{}, we employ an injection methodology that we refer to as \textit{cross-sample} injection.
In this setting, we insert one feature from a \textit{different} sample of the \textit{same benchmark} \eg, a feature taken from another \texttt{CodeSearchNet} instance when in the code summarization task, or a feature taken from another \texttt{BigCodeBench} input when in the code generation task.
For each prompt under test, we first extract its features $\langle f_1, \ldots, f_n \rangle$.
We then compute the average feature length \ie measured as the number of tokens in each feature.
This is used to build a pool of candidate features taken from all \textit{other} prompts in the dataset.
We retain only those candidates whose length ranges within a three-token window around this average.
This ensures that the injected noise is consistent with the form of the original features.
From the resulting pool, we randomly select one candidate feature and inject it at a randomly chosen position in the original feature sequence, leaving the rest of the prompt unchanged.
This setting evaluates \approach's ability to detect contextually inappropriate features, \eg ``\textit{implement binary search}'' in a prompt about string manipulation.

\textbf{Instances Filtering.}
As in \RQ{1}, to ensure that the injected feature is irrelevant, we only evaluate cases in which the cross-sample feature does not affect the model's output.

For each prompt $i$ in the benchmark, we first generate the baseline output $o = M(i)$ using the original prompt.
We then construct its cross-sample variant $ni_{\text{cross}}$ by injecting one feature taken from a different sample of the same benchmark, as described above, and obtain the candidate output $O' = M(ni_{\text{cross}})$.
We retain the pair $\langle ni_{\text{cross}}, M(ni_{\text{cross}}) \rangle$ only if $O' = o$, \ie the injected cross-sample feature does not change the original output.
Consistently with \RQ{1}, we use exact string matching between $o$ and $O'$ to ensure that the injected feature has no impact on the model output, neither from a functional nor from a non-functional perspective, while the other variants are discarded.

\textbf{Metrics.}
We follow the same principles as those used in \RQ{1}.
Consequently, for each retained instance $I$ with injected cross-sample feature $nf$, we compute the \textit{Noise Score} as $S_I = A(nf)$ where $A(nf)$ denotes the attribution assigned by \approach to the injected feature.
The lower values of $S_I$ indicate that the method correctly assigns negligible attribution to the injected feature.

We then analyze the distribution of the noise scores in all filtered instances $\mathcal{D}$ using the mean and median.
To compare \approach against the baselines, we again performed pairwise Wilcoxon signed-rank tests \cite{wilcoxon1992individual}, and computed Cliff's delta effect size \cite{grissom2005effect}. Multiple comparisons are corrected using Holm's correction \cite{holm1979simple}.

\begin{figure}
\footnotesize
\begin{tcolorbox}[title=\textbf{Prompt for LLM-based Feature Attribution}, colback=gray!5, colframe=black,
width=0.8\columnwidth,boxrule=0.15mm]
\texttt{SYSTEM\_PROMPT:} You are a code feature attribution analyst.\\
Your task is to evaluate the importance of individual features in contributing to a given model output.\\
Feature attribution scoring measures how much each feature contributes to the meaning captured in the model output.\\
A higher score (1) indicates that the feature is more essential for understanding what the code does, while a lower score (0) \\
indicates the feature is less relevant or even misleading.\\

\texttt{CRITICAL REQUIREMENTS:}\\
- All attribution scores MUST be between 0.0 and 1.0\\
- The sum of all attribution scores MUST equal exactly 1.0\\
- Provide exactly one score per feature\texttt{"""}\\

\texttt{USER\_PROMPT:}\\Prompt: \texttt{\{prompt\}}\\
Model output: \texttt{\{model\_output\}}\\
Please assign an attribution score to each of the following features: \texttt{\{features\}}\\
Remember that the sum of the feature attribution scores has to be 1.\\
\end{tcolorbox}
\caption{Prompt template for the LLM-as-an-attributor baseline.  Here, \texttt{\{benchmark\_prompt\}} refers to the prompt from the benchmark, \texttt{\{model\_output\}} refers to the output of the model based on the \texttt{\{benchmark\_prompt\}}, and \texttt{\{features\}} refers to the features obtained from the splitter.}
\label{fig:prompt_attributor}

\end{figure}

\textbf{Baselines.}
We compare \approach with two baselines.
First, as in the previous section, we include a \textit{random-based attribution baseline}.
This baseline assigns random values between 0 and 1 to each feature and normalizes them to sum up to 1.
However, in the cross-sample setting of \RQ{2} the random baseline becomes weakly informative.

Inspired by LLM-as-a-judge evaluation paradigms \cite{crupi2025effectiveness,ahmed2025can,weyssow2024codeultrafeedback}, which show that LLMs can approximate human judgments in assessing software artifacts, we adopt an \textit{LLM-as-an-attributor} baseline rather than the random baseline used in \RQ{1}.
This choice is motivated by the fact that cross-sample noise represents a more realistic and challenging setting.
Under these conditions, an \textit{LLM-as-an-attributor} baseline provides a stronger and more realistic comparison, better reflecting a human annotator when judging feature attribution.
In detail, for the \textit{LLM-as-an-attributor} baseline, we use \texttt{GPT-5-mini} with zero-shot prompting to assign attribution scores. We engineered the prompt in \figref{fig:prompt_attributor} that instructs the model to behave as a feature attribution analyst and to distribute importance values between the input features. Specifically, the model must assign a real value score between 0 and 1 to each feature, under the constraint that all scores sum to exactly 1.0, thus producing a normalized importance distribution. This setup allows a direct comparison between the distribution generated by the \textit{LLM-as-an-attributor} and the attribution scores computed by \approach.

\subsection{Experimental Procedure for \RQ{3}}
\begin{figure}[t]
  \centering
  \begin{minipage}[t]{0.49\linewidth}
    \raisebox{0.5\height}{\includegraphics[width=\linewidth]{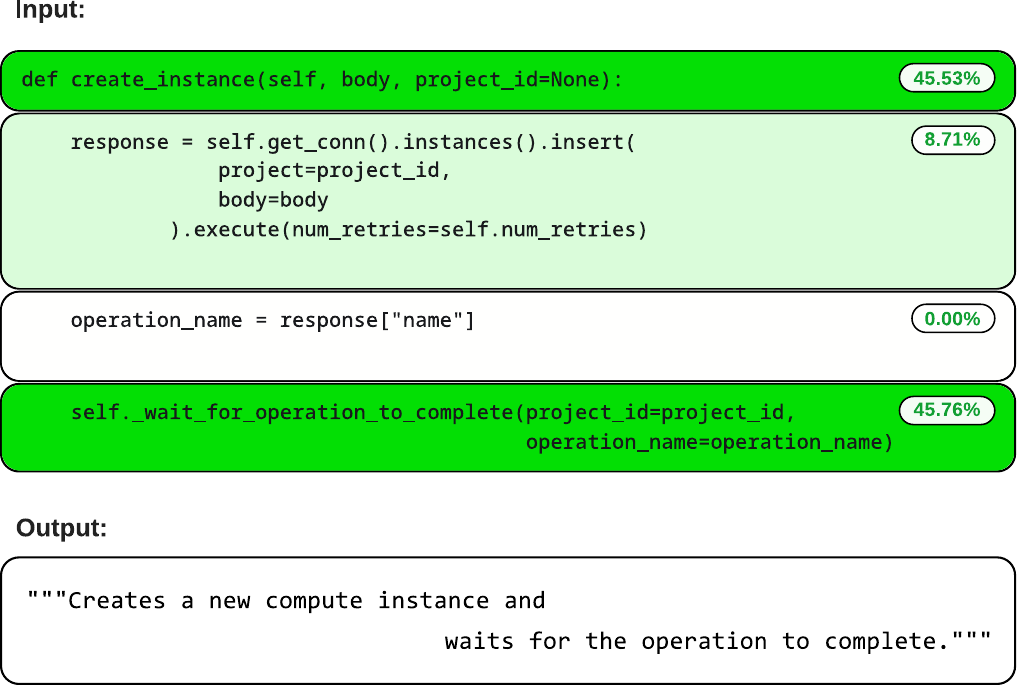}}
  \end{minipage}
  \hfill
  \begin{minipage}[t]{0.50\linewidth}
    \scriptsize
    \begin{tcolorbox}[
      title=\textbf{Survey Questions for Example 1},
      colback=gray!5,
      colframe=black,
      boxrule=0.15mm,
      width=\linewidth
    ]
\textbf{Q1.} Do you agree that the most intensely highlighted parts are also the most important ones behind the model's summary? \textit{(1 = Strongly Disagree, 5 = Strongly Agree)}
\vspace{0.5em}
\begin{center}
\begin{tabular}{@{}c@{\hskip 2.2em}c@{\hskip 2.2em}c@{\hskip 2.2em}c@{\hskip 2.2em}c@{}}
1~\radiobutton & \radiobutton & \radiobutton & \radiobutton & \radiobutton~5 \\
\end{tabular}
\end{center}
\vspace{1.5em}

\textbf{Q2.} If you disagreed (rating <4), what alternative highlighting would you suggest?
\textit{e.g., Feature 1 > Feature 3 > Feature 2}
\vspace{0.5em}
\begin{center}
\rule{6cm}{0.4pt}
\end{center}
\vspace{1.5em}

\textbf{Q3.} To what extent did the highlighting help you understand the model's output? \\
\vspace{-1.5em}
\begin{center}
    \textit{(1 = Not at all, 5 = Completely)}
\end{center}
\vspace{0.5em}
\begin{center}
\begin{tabular}{@{}c@{\hskip 2.2em}c@{\hskip 2.2em}c@{\hskip 2.2em}c@{\hskip 2.2em}c@{}}
1~\radiobutton & \radiobutton & \radiobutton & \radiobutton & \radiobutton~5 \\
\end{tabular}
\end{center}
\vspace{1.5em}

\textbf{Q4.} Did the highlighting help determine if the summary aligns with the input code?
\vspace{0.5em}
\begin{center}
\begin{tabular}{@{}c@{\hskip 2.2em}c@{\hskip 2.2em}c@{\hskip 2.2em}c@{\hskip 2.2em}c@{}}
1~\radiobutton & \radiobutton & \radiobutton & \radiobutton & \radiobutton~5 \\
\end{tabular}
\end{center}
\vspace{1.5em}

\textbf{Q5.} Would the highlighting assist you in accepting, modifying, or rejecting the generated summary?
\vspace{0.5em}
\begin{center}
\begin{tabular}{@{}c@{\hskip 2.2em}c@{\hskip 2.2em}c@{\hskip 2.2em}c@{\hskip 2.2em}c@{}}
1~\radiobutton & \radiobutton & \radiobutton & \radiobutton & \radiobutton~5 \\
\end{tabular}
\end{center}
\vspace{1.5em}

\textbf{Q6.} Please motivate your previous answer.
\vspace{0.5em}
\begin{center}
\rule{6cm}{0.4pt}
\end{center}
\vspace{1.5em}

\textbf{Q7.} Other comments/notes.
\vspace{0.5em}
\begin{center}
\rule{6cm}{0.4pt}
\end{center}

    \end{tcolorbox}
  \end{minipage}
  \caption{An example of a code summarization \approach attribution from our survey study.}
  \label{fig:survey-data}
\end{figure}
To answer \RQ{3} and evaluate the practical effectiveness of \approach, we conducted a survey with \nparticipants practitioners. Each participant was asked to evaluate six attribution examples generated by \approach in the two tasks under study: three from code generation and three from code summarization. This practitioner-focused evaluation is essential because assigning low attribution scores to irrelevant features is a necessary but not sufficient condition to conclude that the assigned scores are correct. Indeed, such a preliminary evaluation does not reveal whether the explanations are actually \emph{useful} to humans. Previous work by Li \etal \cite{li2024machines} highlights this gap, showing that developers often perceive existing explanation methods as misaligned with their reasoning, raising concerns about practical adoption. Our survey directly addresses this issue by asking practitioners to judge whether \approach's feature-level attributions align with their own understanding and provide actionable support for decision-making.

Building on this motivation, we designed the survey to cover a diverse set of examples that would allow participants to judge the clarity and usefulness of the attributions. The central part of the survey consisted of an evaluation of the explanations generated by \approach on 3 examples of code summarization and 3 code generation. To select the examples, we used the following procedure. First, we computed descriptive statistics in terms of the size (number of features) of the instances in the benchmarks we considered.
Then, for each dataset, we selected three representative samples according to length: one at the first quartile, one at the median, and one at the third quartile of the length distribution.
For each selected instance, we run GPT-4.1-nano to perform the task (either code generation or summarization) and used \approach to generate the explanation by assigning attribution scores to each feature. We chose GPT-4.1-nano because it is one of the selectable models in GitHub Copilot, making it representative of configurations accessible to developers. Note again that we are not interested in task accuracy but rather in the quality of the explanations.

For each example, participants were shown an input-output pair, along with the corresponding \approach attributions to the input features in percentage form. These attributions were also visualized using color-coded highlights, where the intensity of the color indicated the estimated contribution of each feature to the model's output. Participants were asked to evaluate the \textit{relevance} and \textit{clarity} of such attributions using a combination of Likert-scale ratings \cite{likert1932technique} and open-ended responses.
The evaluation questions were designed following the framework proposed by Marcinkevics \etal \cite{marcinkevivcs2020interpretability}, which distinguishes between two complementary dimensions: (i) \textit{Faithfulness:} Do the attributions accurately reflect the true influence of each input feature on the model's output? We investigated this by asking participants whether the most highly attributed features aligned with their own judgment of importance; (ii) \textit{Usefulness:} Are the attributions understandable, actionable, and supportive of real-world development tasks such as reviewing, accepting, or modifying model outputs? This was evaluated by assessing whether the attributions aligned with the task requirements and whether the participants perceived them as helpful for decision-making. 

Each participant reviewed all six examples and provided feedback addressing both faithfulness and usefulness. \figref{fig:survey-data} illustrates one of the examples shown to participants when evaluating code summarization \approach attribution. Each example included both the model's input (\eg the function to be summarized), the output generated by the LLM, and the corresponding attributions produced by \approach, visualized as feature-level highlights. Participants were asked to evaluate the \textit{faithfulness}--captured by \textbf{Q1}--(\figref{fig:survey-data}), and \textit{usefulness} that is mapped onto the remaining questions (\textbf{Q3-Q5}).

We present both quantitative and qualitative findings derived from participant responses (\secref{sec:results}). Quantitatively, we analyze the distribution of Likert-scale responses across six examples and visualize them using box plots to identify trends in perceived faithfulness and usefulness. We also report qualitative feedback from open-ended responses, highlighting insights from participants found particularly informative, such as alignment with model behavior and areas for improvement. Together, these findings offer a nuanced view of how \approach is perceived in practical SE tasks.

\section{Results}
\label{sec:results}

In this section, we first report our findings on \approach's ability to assign low attribution to injected noise, starting with the preliminary check performed in \RQ{1} and followed by the more realistic setting performed in \RQ{2}. We then present the results of our practitioner-focused user study in \RQ{3}, examining the faithfulness and usefulness of the explanations produced by \approach.

\subsection{\RQ{1}: Random Noise Attribution Scores}
In \tabref{tab:rq1_res} we present our experimental results for \RQ{1}. These are organized by task (Code Generation and Code Summarization) and models (\texttt{Qwen2.5-Coder-\{3,7,32\}B} and \texttt{GPT-4.1-Nano}). We compare \approach against a \textit{Random-baseline} under the \textit{nonsensical} injection setting.
Each entry reports: (i) the \textit{Noise Score}: computed per instance as the attribution assigned to the injected noise feature, summarized using the Mean and Median across all instances, (ii) \textit{\#Samples}: instances retained after applying filtering criteria from \secref{sec:design}, (iii) the statistical comparison \textit{vs \approach}, reported through the adjusted \textit{$p$-value} from the Wilcoxon signed-rank test and Cliff's \textit{$\delta$} effect size with its corresponding magnitude \ie (N)egligible, (S)mall, (M)edium, or (L)arge.
We highlight in bold the attributor and the relative \textit{Noise Score} values that outperform each other according to statistical tests ($p<0.05$). If no statistically significant difference is observed ($p\geq0.05$), the values are reported in regular font.

\begin{table}[t]
\centering
\caption{Comparison of \approach and \textit{Random-baseline} for code generation and code summarization tasks. Lower is better.}
\label{tab:rq1_res}
\resizebox{0.6\columnwidth}{!}{%
\begin{tabular}{lll ccccc}
\toprule
\multirow{3}{*}{\textbf{Task}} & \multirow{3}{*}{\textbf{Model}} & \multirow{3}{*}{\textbf{Attributor}} &
  \multicolumn{5}{c}{\textbf{Nonsensical}} \\
\cmidrule(lr){4-8}
 &  &  &
  \multicolumn{2}{c}{\textbf{Noise Score}} & \multirow{2}{*}{\textbf{\#Samples}} & \multicolumn{2}{c}{\textbf{vs \approach}} \\
\cmidrule(lr){4-5} \cmidrule{7-8}
 &  &  & \textbf{Mean} & \textbf{Median} &  & $\bm{p}$\textbf{-value} & $\bm{\delta}$ \\
\midrule
% --- BCB ---
\multirow{8}{*}{\rotatebox[origin=c]{90}{Code Generation}}

 & \multirow{2}{*}{\texttt{\texttt{\texttt{QC-3B}}}}
 & \textbf{\approach} & \textbf{0.011} & \textbf{0.000} & \multirow{2}{*}{245} &           -- &                -- \\
 & & \textit{Random}           & 0.168 &          0.151 &                      & $\bm{<0.05}$ & \textbf{-0.9 (L)} \\
\cmidrule{2-8}

 & \multirow{2}{*}{\texttt{\texttt{QC-7B}}}
 & \textbf{\approach} & \textbf{0.021} & \textbf{0.000} & \multirow{2}{*}{395} &           -- &                -- \\
 & & \textit{Random}  &          0.161 &          0.140 &                      & $\bm{<0.05}$ & \textbf{-0.9 (L)} \\
\cmidrule(lr){2-8}

 & \multirow{2}{*}{\texttt{\texttt{QC-32B}}}
 & \textbf{\approach} & \textbf{0.010} & \textbf{0.000} & \multirow{2}{*}{341} &           -- &                -- \\
 & & \textit{Random}  &          0.167 &          0.158 &                      & $\bm{<0.05}$ & \textbf{-0.9 (L)} \\
\cmidrule(lr){2-8}

 & \multirow{2}{*}{\texttt{\texttt{GPT-4.1-Nano}}}
 & \textbf{\approach} & \textbf{0.029} & \textbf{0.000} & \multirow{2}{*}{127} &           -- &                -- \\
 & & \textit{Random}  &          0.180 &          0.164 &                      & $\bm{<0.05}$ & \textbf{-0.8 (L)} \\
\midrule

% --- CSN ---
\multirow{8}{*}{\rotatebox[origin=c]{90}{\texttt{Code Summarization}}}

 & \multirow{2}{*}{\texttt{\texttt{\texttt{QC-3B}}}}
 & \textbf{\approach} & \textbf{0.049} & \textbf{0.000} & \multirow{2}{*}{159} &           -- &                -- \\
 & & \textit{Random}  &          0.173 &          0.154 &                      & $\bm{<0.05}$ & \textbf{-0.7 (L)} \\
\cmidrule(lr){2-8}

 & \multirow{2}{*}{\texttt{\texttt{QC-7B}}}
 & \textbf{\approach} & \textbf{0.021} & \textbf{0.000} & \multirow{2}{*}{284} &           -- &                -- \\
 & & \textit{Random}  &          0.194 &          0.170 &                      & $\bm{<0.05}$ & \textbf{-0.9 (L)} \\
\cmidrule(lr){2-8}

 & \multirow{2}{*}{\texttt{\texttt{QC-32B}}}
 & \textbf{\approach} & \textbf{0.017} & \textbf{0.000} & \multirow{2}{*}{203} &           -- &                -- \\
 & & \textit{Random}  &          0.194 &          0.171 &                      & $\bm{<0.05}$ & \textbf{-0.9 (L)} \\
\cmidrule(lr){2-8}

 & \multirow{2}{*}{\texttt{\texttt{GPT-4.1-Nano}}}
 & \textbf{\approach} & \textbf{0.034} & \textbf{0.000} & \multirow{2}{*}{217} &           -- &                -- \\
 & & \textit{Random}  &          0.181 &          0.152 &                      & $\bm{<0.05}$ & \textbf{-0.8 (L)} \\

\bottomrule
\end{tabular}%
}
\end{table}

\textbf{On Code Generation.}
For the code generation task, \approach always produces lower \textit{Noise Scores} than the \textit{Random-baseline}.
The medians for \approach are $0$ across all models, indicating that in at least half of the instances zero attribution is assigned to the injected noise.
\approach's means are an order of magnitude smaller than random. For example, on \texttt{Qwen2.5-Coder-3B}, \approach achieves a Noise Score Mean of $0.011$ in contrast of the one of the \textit{Random-baseline} of $0.168$.
Similar differences appear on 7B and 32B ($\sim\!17\times$ lower on 32B), and on \texttt{GPT-4.1-Nano} ($\sim\!6\times$ lower).

\textbf{On Code Summarization.}
For the code summarization task we observe the same patterns.
The medians for \approach are $0$ for all models, and the means are generally $4$–$11\times$ lower than \textit{Random-baseline}.
For example, on \texttt{Qwen2.5-Coder-7B} \approach produces a Noise Score Mean of $0.021$ against the $0.194$ of the \textit{Random-baseline} ($\sim\!9\times$ lower) and on \texttt{Qwen2.5-Coder-32B} $0.017$ vs $0.194$ ($\sim\!11\times$ lower).

In both datasets and all models, all pairwise comparisons are statistically significant ($p<0.05$) with large effect sizes ($|\delta|=0.7$–$0.9$), confirming that \approach reliably assigns negligible attribution to irrelevant features. 

\mycolorbox{1}{\approach consistently outperforms the \textit{Random-baseline} across tasks and models.}

\subsection{\RQ{2}: Cross-sample Noise Attribution Scores}
\begin{table}[t]
\centering
\caption{Comparison between \approach, \textit{Random-baseline}, and \textit{LLM-as-an-attributor} on code generation and code summarization tasks. Lower is better.}
\label{tab:llm_attributor_compare}
\resizebox{0.7\columnwidth}{!}{%
\begin{tabular}{c l l ccccc}
\toprule
\multirow{3}{*}{\textbf{Task}} & \multirow{3}{*}{\textbf{Model}} & \multirow{3}{*}{\textbf{Attributor}} &
  \multicolumn{5}{c}{\textbf{Cross-sample}} \\
\cmidrule(lr){4-8}
 &  &  &
  \multicolumn{2}{c}{\textbf{Noise Score}} & \multirow{2}{*}{\textbf{\#Samples}} & \multicolumn{2}{c}{\textbf{vs \approach}} \\
\cmidrule(lr){4-5}\cmidrule{7-8}
 &  &  & \textbf{Mean} & \textbf{Median} &  &  $\bm{p}$\textbf{-value} & $\bm{\delta}$  \\
\midrule
% --- BCB ---
\multirow{12}{*}{\rotatebox[origin=c]{90}{Code Generation}}

 & \multirow{3}{*}{\texttt{\texttt{\texttt{QC-3B}}}}
 & \approach                                   &          0.009 &          0.000 &     &        --    &                -- \\
 & & \textit{LLM-as-an-attributor}             &          0.028 &          0.020 & 114 & $\bm{<0.05}$ & \textbf{-0.5 (L)} \\
 & & \textit{Random}                           &          0.157 &          0.138 &     & $\bm{<0.05}$ & \textbf{-0.9 (L)} \\
\cmidrule(lr){2-8}

 & \multirow{3}{*}{\texttt{\texttt{QC-7B}}}
 & \approach                                   &          0.018 &          0.000 &     &        --    &                -- \\
 & & \textit{LLM-as-an-attributor}             &          0.022 &          0.010 & 210 & $\bm{<0.05}$ & \textbf{-0.4 (M)} \\
 & & \textit{Random}                           &          0.170 &          0.155 &     & $\bm{<0.05}$ & \textbf{-0.9 (L)} \\
\cmidrule(lr){2-8}

 & \multirow{3}{*}{\texttt{\texttt{QC-32B}}}
 & \approach                                   &          0.018 &          0.000 &     &        --    &                -- \\
 & & \textit{LLM-as-an-attributor}             &          0.025 &          0.010 & 145 & $\bm{<0.05}$ & \textbf{-0.4 (M)} \\
 & & \textit{Random}                           &          0.146 &          0.127 &     & $\bm{<0.05}$ & \textbf{-0.9 (L)} \\

\cmidrule(lr){2-8}

 & \multirow{3}{*}{\texttt{\texttt{GPT-4.1-Nano}}}
 & \approach                                   &          0.025 &          0.000 &    &        --    &                -- \\
 & & \textit{LLM-as-an-attributor}             &          0.014 &          0.000 & 54 &         0.868 &          -0.1 (N) \\
 & & \textit{Random}                           &          0.163 &          0.162 &    &  $\bm{<0.05}$ & \textbf{-0.8 (L)} \\

\midrule

% --- CSN ---
\multirow{12}{*}{\rotatebox[origin=c]{90}{Code Summarization}}

 & \multirow{3}{*}{\texttt{\texttt{\texttt{QC-3B}}}}
 & \approach                                   &          0.034 &          0.000 &     &        --    &                -- \\
 & & \textit{LLM-as-an-attributor}             &          0.025 &          0.020 & 119 &       0.133  &          -0.4 (M) \\
 & & \textit{Random}                           &          0.163 &          0.151 &     & $\bm{<0.05}$ & \textbf{-0.8 (L)} \\
\cmidrule(lr){2-8}

 & \multirow{3}{*}{\texttt{\texttt{QC-7B}}}
 & \approach                                   &          0.019 &          0.000 &     &        --    &                -- \\
 & & \textit{LLM-as-an-attributor}             &          0.026 &          0.020 & 186 & $\bm{<0.05}$ & \textbf{-0.6 (L)} \\
 & & \textit{Random}                           &          0.180 &          0.143 &     & $\bm{<0.05}$ & \textbf{-0.9 (L)} \\
\cmidrule(lr){2-8}

 & \multirow{3}{*}{\texttt{\texttt{QC-32B}}}
 & \approach                                   &          0.017 &          0.000 &     &        --    &                -- \\
 & & \textit{LLM-as-an-attributor}             &          0.022 &          0.020 & 143 & $\bm{<0.05}$ & \textbf{-0.6 (L)} \\
 & & \textit{Random}                           &          0.166 &          0.148 &     & $\bm{<0.05}$ & \textbf{-0.9 (L)} \\
\cmidrule(lr){2-8}

 & \multirow{3}{*}{\texttt{\texttt{GPT-4.1-Nano}}}
 & \approach                                   &          0.035 &          0.000 &     &        --    &                -- \\
 & & \textit{LLM-as-an-attributor}             &          0.031 &          0.020 & 148 & $\bm{<0.05}$ & \textbf{-0.4 (M)} \\
 & & \textit{Random}                           &          0.190 &          0.177 &     & $\bm{<0.05}$ & \textbf{-0.8 (L)} \\
\bottomrule
\end{tabular}%
}
\end{table}

In \tabref{tab:llm_attributor_compare} we shows the results of the comparison of \approach with both \textit{Random-baseline} and \textit{LLM-as-an-attributor} on the more realistic setting.
The results are organized as in \tabref{tab:rq1_res}.

\begin{figure}[t]
  \centering
  \includegraphics[width=0.6\linewidth]{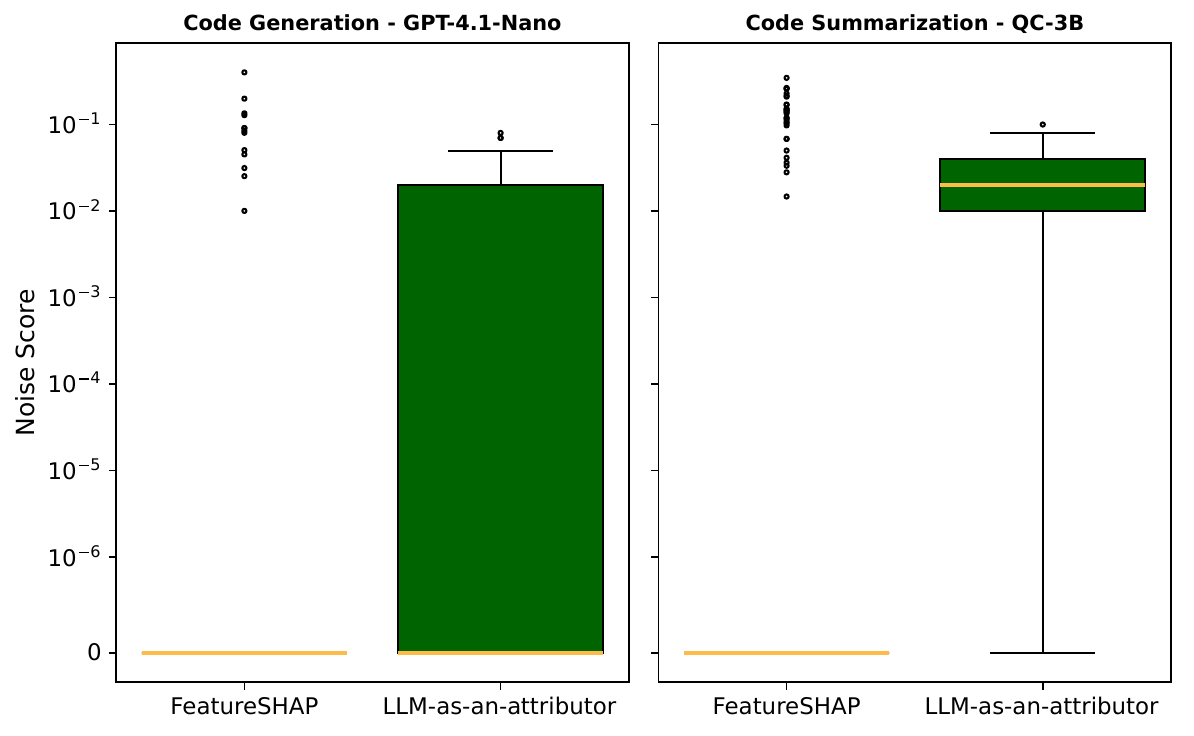}
  \caption{On the left, \textit{Noise Score} distributions of \approach and \textit{LLM-as-an-attributor} on \texttt{GPT-4.1-Nano} for \texttt{BigCodeBench}. On the right, \textit{Noise Score} distributions of \approach and \textit{LLM-as-an-attributor} on \texttt{Qwen2.5-Coder-3B-Instruct} for \texttt{CodeSearchNet}. Lower is better.}
  \label{fig:bcb_and_csn_boxplots_not_significant}
\end{figure}

\textbf{On Code Generation.}
On code generation, \approach achieves consistently lower means than both \textit{Random-baseline} and \textit{LLM-as-an-attributor} on all \texttt{Qwen2.5-Coder} variants, with statistically significant differences ($p<0.05$) and medium-to-large effect sizes, \ie $\delta=-0.8$ to $-0.9$ against \textit{Random-baseline}, and $\delta=-0.4$ to $-0.5$ against \textit{LLM-as-an-attributor}.
Medians for \approach are $0$ across all models, while \textit{LLM-as-an-attributor} shows non-zero medians (0.01–0.02), and \textit{Random-baseline} between 0.127-0.177. This shows that \approach generally assigns \emph{no} attribution to the injected noise on at least half of the instances.
For \texttt{GPT-4.1-Nano}, \textit{Random-baseline} produces higher mean and median (0.163, and 0.162), while \textit{LLM-as-an-attributor} has a slightly lower mean (0.014 vs 0.025) than \approach and both medians are $0$; the difference is not significant ($p=0.868$, $\delta=-0.1$, negligible).
However, looking at the left side of \figref{fig:bcb_and_csn_boxplots_not_significant}, we can observe that \approach distribution is closely concentrated around zero compared to the \textit{LLM-as-an-attributor}, which exhibits a greater variability.

\textbf{On Code Summarization.}
On code summarization, results are mixed but generally in favor of \approach. Generally, \approach consistently assigns lower attributions to noise with respect to the \textit{Random-baseline} with statistically significant differences and large effect size ($\delta=-0.9$).
Furthermore, for \texttt{Qwen2.5-Coder-7B/32B}, \approach produces lower means than the \textit{LLM-as-an-attributor} (0.019/0.017 vs 0.026/0.022) with statistically significant differences ($p<0.05$) and large effects ($\delta=-0.6$).
On the other hand, for \texttt{Qwen2.5-Coder-3B}, the mean difference (0.034 vs 0.025) is not significant ($p=0.133$), while the negative $\delta$ shows a tendency towards lower per-instance scores for \approach despite a few larger outliers increasing its mean.
This pattern is visible in the right panel of \figref{fig:bcb_and_csn_boxplots_not_significant}. \approach's median are set to $0$ with a compact interquartile range, while \textit{LLM-as-an-attributor} shows a higher median and broader spread.
For \texttt{GPT-4.1-Nano}, the mean for \approach is slightly higher (0.035 vs\ 0.031), yet the Wilcoxon test favors \approach ($p<0.05$, $\delta=-0.4$, medium), which, together with the medians (0.000 vs\ 0.020), indicates that \approach more often assigns near-zero attribution but occasionally produces higher attributions.

Overall, across datasets and models, \approach always outperforms the \textit{Random-baseline} and either matches or surpasses \textit{LLM-as-an-attributor}.
When differences are significant, they favor \approach with medium to large effects. When not significant, distributions still show \approach concentrated at zero (median $=0$), whereas \textit{LLM-as-an-attributor} more frequently assigns non-negligible attribution to noise (\eg medians of $0.01$–$0.02$).

Importantly, however, \approach offers greater transparency than \textit{LLM-as-an-attributor}.
\approach SHAP-based attributions process is fully visible to users, instead, the \textit{LLM-as-an-attributor} works as a fully black-box system providing only final scores.
In addition, the \textit{LLM-as-an-attributor} inherently leverages its own understanding of text (\ie natural language and code) to inspect the \textit{content} of each feature. This allows it to detect injected noise directly from its semantic incongruity with the surrounding prompt, rather than from a principled assessment of the model's actual dependency on a feature. In contrast, \approach computes the feature attributions strictly from systematic perturbations and model responses, ensuring that attributions genuinely reflect the model's own behavior.

\mycolorbox{2}{\approach outperforms the \textit{Random-baseline}, and performs on par with or better than the \textit{LLM-as-an-attributor}.}

\subsection{\RQ{3}: \approach Alignment with Human Perception.}
\begin{wrapfigure}{r}{0.33\linewidth}
    \centering
    \vspace{-10pt}
    \includegraphics[width=\linewidth]{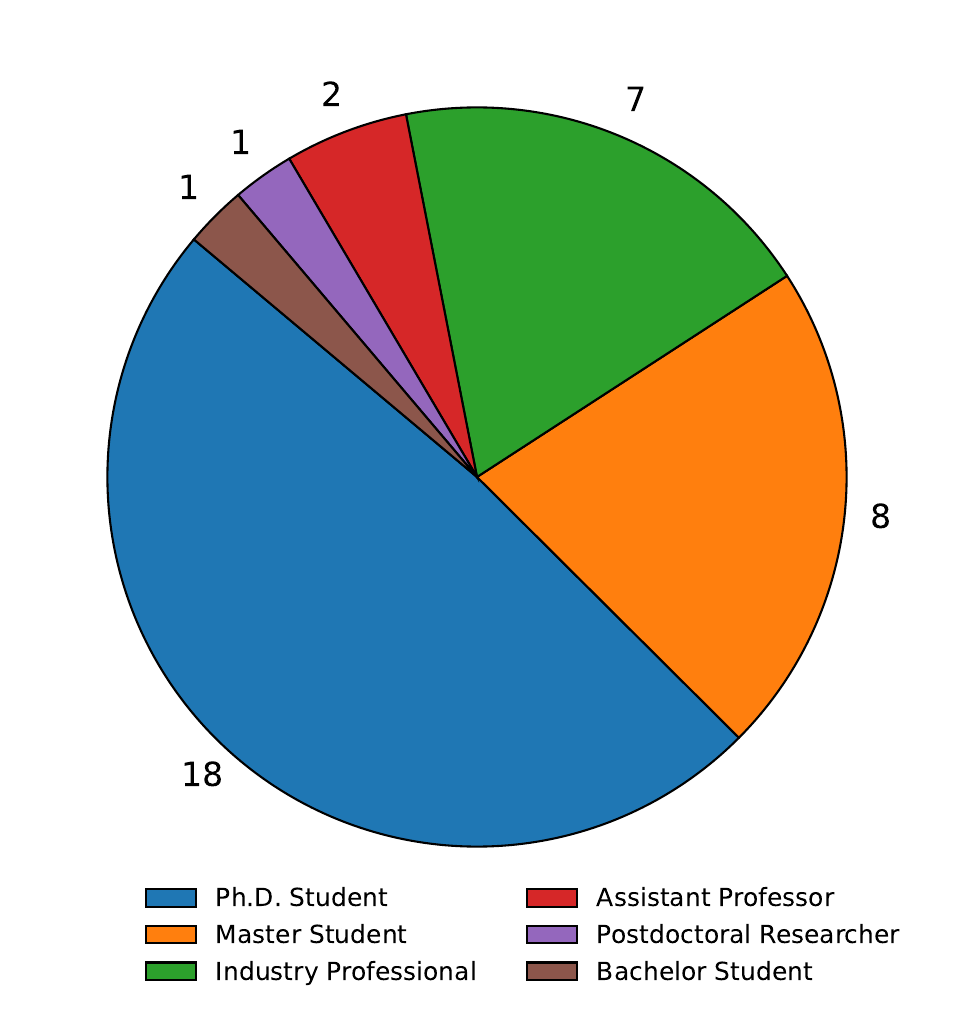}
    \caption{Participants job positions.}
    \label{fig:res_rq2_}
\end{wrapfigure}

In \figref{fig:res_rq2_} we provide an overview of the background of the survey participants, in which we show the proportions of participants belonging to five professional categories: Ph.D. students, master's students, industry professionals, assistant professors, and postdoctoral researchers. Most of them (48.6\% ) are Ph.D. students, while the 21.6\% of the participants involved are Master students, and 18.9\% Industry Professionals (\eg software engineers, data scientist, \etc).

\begin{figure}[t]
  \centering
  \includegraphics[width=\linewidth]{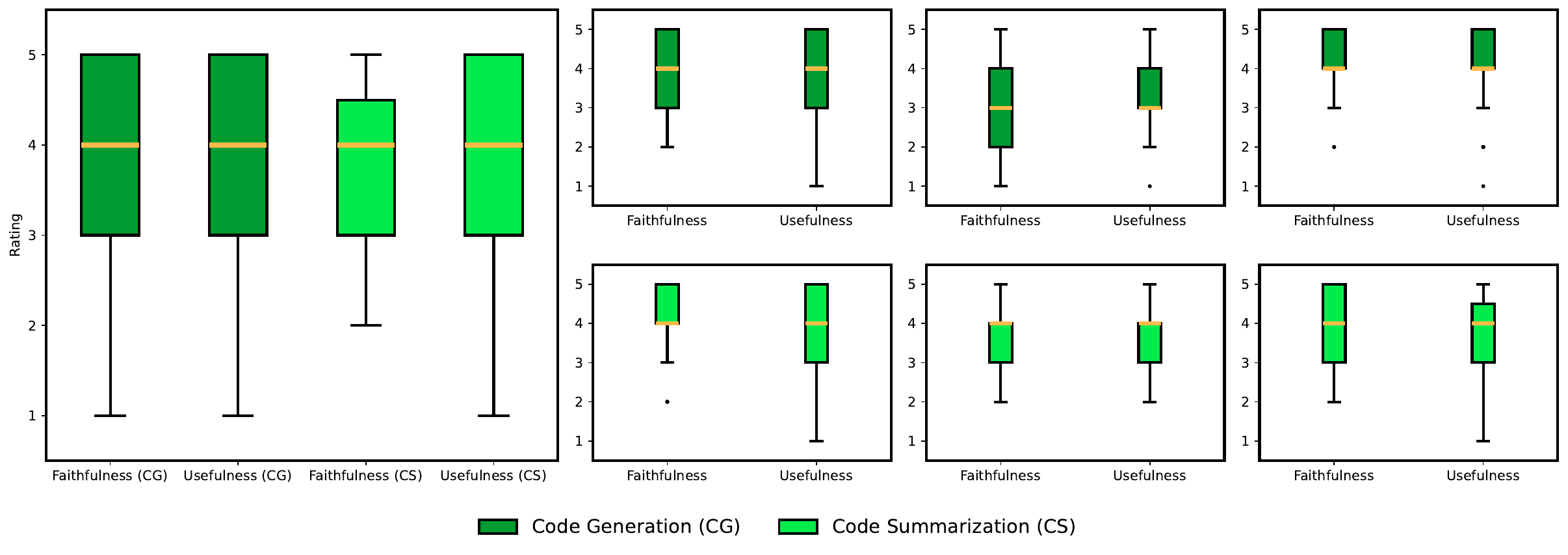}
  \caption{Participant ratings of \approach for faithfulness and usefulness across code generation (CG) and code summarization (CS). The left panel provides an overall summary per task. The right panels show detailed results for individual examples, top row for code generation and bottom row for code summarization.}
  \label{fig:res_rq2}
\end{figure}

We report in \figref{fig:res_rq2} the distribution of Likert-scale ratings for responses related to \textit{faithfulness} and \textit{usefulness}, as provided by participants in both code generation and summarization examples of the survey.
The left panel summarizes the overall ratings for both code generation (CG) and code summarization (CS) tasks, aggregating all evaluated examples per task. The right panel shows the detailed distributions for each individual example. The upper row corresponds to code generation examples, and the bottom row reports results for code summarization ones.

\textbf{Quantitative Analysis.}
Participant ratings indicate that \approach produces explanations perceived as both \textit{faithful} and \textit{useful} across the evaluated tasks.
As shown in \figref{fig:res_rq2}, the distributions of Likert-scale responses for both dimensions tend consistently toward higher values for both tasks, with medians equal to 4 on the 5-point scale.
The interquartile ranges span mostly between 3 and 5, showing moderate consensus among participants. This represent that feature-level attributions generated by \approach accurately reflect the model's decision process and are practically helpful, according to the participants.
The overall boxplots (left panel in \figref{fig:res_rq2}) show that ratings for code generation (CG) and code summarization (CS) are similar, suggesting that \approach generalizes well across distinct SE tasks involving both natural and programming languages.
However, while the distribution for code generation shows slightly wider variability, particularly for \textit{usefulness} with respect to the code summarization counterpart, the median and upper quartiles remain high, confirming that explanations remain consistently well aligned with participants reasoning despite the greater inherent complexity of code generation compared to code summarization.

Analyzing the ratings at individual example level (right part of \figref{fig:res_rq2}), we observe a similar pattern across all six examples.
In detail, for the first code generation example, participants provided uniformly high ratings for both \textit{faithfulness} and \textit{usefulness}, although a more variability is noticed for \textit{usefulness} related ratings.
Instead, in the second code generation example, the median rating for \textit{faithfulness} is equal to 3 showing moderate participants alignment with \approach attribution scores. 
For the third code generation example, both \textit{faithfulness} and \textit{usefulness} receives higher ratings, with medians at 4 and upper quartiles extending to 5. The limited variance observed reports that participants consistently perceived \approach's explanations as indicative of the factors influencing the model's generation behavior, and as helpful to be leveraged in practice.
Moving to the code summarization task, the first example shows that both \textit{faithfulness} and \textit{usefulness} ratings centered around 4 with strong upper quartile for \textit{faithfulness} ratings.
In the second summarization example, a slightly wider variability appears in the \textit{faithfulness} ratings, with a median value equal to 4.
Finally, for the third summarization example, both \textit{faithfulness} and \textit{usefulness} achieve a median equal to 4, with upper quartiles at 5 and a few low outliers.

In general, for both code generation and code summarization, \textit{faithfulness} median ratings often equal to 4 with the upper quartiles that reaches 5 for the first and third example, while for the second example the median is equal to 3.
This consistency indicates that participants largely agreed that the most important features according to \approach indeed correspond to the parts of the input that mostly influenced the model to infer the output.
\textit{Usefulness} ratings follow a similar trend, with median values equal to 4 for all the examples except for the second code generation and code summarization examples.
Importantly, no significant difference emerges between \textit{faithfulness} and \textit{usefulness} ratings within the same example, implying that participants found faithful explanations also directly useful for their assessment of model behavior.

\textbf{Qualitative Analysis.}
Looking at the qualitative feedback reported by participants, we observe consistent evidence that \approach provides both \textit{faithful} and \textit{useful} explanations across the evaluated tasks. 
In the code generation examples, participants emphasized the usefulness of the highlighting distribution (\ie feature attributions) as guidance to understand the model behavior and verify requirement coverage.
For example, participants stated \textit{"The highlighting makes it super easy to see what kind of code I'm trying to generate and what the AI focused on"}, \textit{"Highlighting helps to focus on the important parts that are integral for the code"}, or \textit{"The highlighting clearly maps requirements to code sections ..."}.
This successfully aligns with our conjecture \ie splitting in semantic features rather than in single tokens aligns better with practitioners point of view.
Others confirmed this aspect more explicitly, noting that \textit{"The highlighting distribution provides a clear indication of which parts of the code are relevant to the requirements ..."} and \textit{"Highlighting the input according to its degree of influence is useful for focusing attention on the most relevant parts of the request"}. 
These comments show how participants perceived \approach useful, providing transparency for what the model prioritized when producing the output.
A minority of participants pointed out limitations, mainly related to imbalances in the attribution scores. 
Some reported that \textit{"Highlighting is helpful but sometimes misses key requirements ..."}. This kind of comment reveals occasional mismatches between human perception and \approach attributions. Instead, comments as the following, \textit{"... I might mistakenly accept generated code that does the highlighting things well and ignore the non-highlighted things ..."}, show that while participants appreciate feature attributions, they manually must cross-validate the code to ensure completeness. In fact, our goal is not to encourage developers to automatically accept generated outputs based on the explanations, but rather to \textit{support} their reasoning process to promote transparency and awareness of the model generation process. For example, this can enable practitioners to spot unreliable attributions and to critically assess when to trust the model or when to refine the prompt or the code itself. In this sense, explanations serve as decision aids, not as decision replacements.
Some of the participants indeed align with our assumption stating that \textit{"... seeing that the model puts more emphasis in the initialization and the middle function rather than the regex code will make me re-prompt it better"}.
Nevertheless, the general trend remains positive, with several participants stating that \textit{"It indeed helps to understand whether the model tackled all the requirements in the prompt"}.

Similar patterns emerge for the code summarization task. 
Participants highlighted the benefit of traceability between the source code and the generated documentation, describing that \textit{"It helps me quickly identify the parts of the code that are considered by the model, and in this specific case that the parts that are given the greatest consideration are precisely the important ones."} and \textit{"This is a very obvious example to see precisely where the model drew its output from. The two parts of the output directly correspond to the two most highlighted parts of the code"}. 
Such comments suggest that participants found the explanations faithful, and helpful for verifying the consistency between code and the generated summaries. 
Some participant disagreed, mentioning that \textit{"... the highlighting does not accurately reflect the importance of each feature in the summary"}, indicating that small inconsistencies in attributions intensity could sometimes reduce perceived faithfulness.

In addition, some also suggested newer perspectives and future directions for more human-aligned explanations. For example, the following comment, \textit{"It would be even easier if the text, or the individual feature, were associated with the piece of code"}, suggests for a specialization foreseeing a mapping between feature attributions and output features.
One other participant also wondered about a multiple-steps explanation process in which first a more general overview is provided, followed by the possibility to progressively explore finer-grained details of the attribution, \ie \textit{"... it might be interesting to see the LLM work with more granular information ..."}.

In summary, the qualitative feedback confirms the quantitative findings. Participants perceived \approach as both \textit{faithful} and \textit{useful}, improving the understanding of the model generation process and supporting decision-making in results inspection. 
Even when misalignment emerged, \approach attributions enhanced transparency rather than undermining trust, as participants used such discrepancies to critically assess where the model focus diverged from their own expectations.

\mycolorbox{3}{
Participants rate \approach's explanations as both faithful and useful across both code generation and code summarization tasks. Qualitative feedback further shows that feature-level attributions help users understand and check the model's behavior.
}

\section{Implications of Our Findings and Future Work}
\label{sec:implications}

Our study presents strong evidence that \approach offers a practical, scalable, and domain-aligned solution to the challenge of explainability in software engineering automation. Based on our findings, we define the following implications for researchers, practitioners, and tool builders.

\paragraph{\scalebox{0.8}{\faBook} \xspace \textbf{\textit{Researchers.}}}
Our findings emphasize the importance of domain-adapted explainability techniques that go beyond token-level reasoning by leveraging semantically-meaningful features that reflect the way developers naturally interpret and reason about code.

\approach opens new avenues for exploring feature-based attribution in a wider range of tasks, including code-to-code scenarios such as automated program repair, refactoring, and code translation.
Indeed, the modular design of \approach also allows researchers to explore and specialize the different components (\ie splitter, modifier, and comparator) for a specific task at hand. For example, regarding the automated program repair (APR) task, the splitter can be adapted to isolate the buggy statement or its corresponding AST subtree as a dedicated feature, while grouping the remaining statements into features of comparable granularity to preserve the structural balance of the code. The modifier can then replace each feature with semantics-preserving alternatives, such as substituting the buggy statement with a minimal repair candidate or a paraphrased equivalent, thus maintaining syntactic validity while probing the model's dependence on that section. Finally, the comparator can leverage code-aware similarity metrics (\eg CodeBLEU) to quantify how each substitution alters the patch produced by the model. This could enable \approach to show whether the repair ability of the model truly depends on the buggy section or is influenced by an unrelated context, providing fine-grained insight into the behavior and reliability of the LLM repair process.
In general, while initial feature splitting rules may require domain expertise, we have shown that this step can be effectively automated using LLMs or other techniques. This demonstrates that the framework is not only extensible but also scalable, enabling its application in diverse tasks where LLMs explainability is essential.

In addition, \approach also lends itself to \textit{systematic} analysis of LLMs behavior.
Since attributions are computed in a model-agnostic and fully automated way, one can run \approach over large batches of generations (\eg across prompts) to spot recurring failures, spurious correlations, and contrast attribution before and after fine-tuning.
For example, given a large dataset of prompts that share a common structure, researchers can identify recurring prompt components (\eg initial instructions, requirements, constraints, or non-functional specifications) and apply \approach to obtain feature-level attributions for each instance.
Since the features would correspond to known prompt sections, aggregating these for each instance can show which parts of the prompt the model consistently relies on, enabling the analysis and exploration of the model global behavior.
Similarly, running \approach on pairs of models, such as a model tuned to instruction and its fine-tuned counterpart, on the same dataset can allow researchers to directly contrast their attributions on the same inputs.
Such comparisons can let one study how fine-tuning shifts the model's dependence on specific features, studying why behavior changes and which modifications could have altered model reasoning.

Furthermore, it could be meaningful to connect \approach to the mechanistic interpretability \cite{bereska_mechanistic_2024} through \textit{causal interventions}. In practice, this means testing whether removing or altering specific input features (input-level ablations) or modifying their internal activations (activation patching) leads to predictable changes in the model's output. Such tests act as a causal sanity check that high-attribution features truly \emph{matter} for the model's behavior. At the same time, it could extend \approach with \textit{dependency-aware attribution}, which explicitly accounts for the fact that code features are often interdependent. By estimating interaction effects and using AST-aware masking to preserve syntactic and semantic validity, this enhancement can reveal whether the importance of a feature (\eg a docstring) comes from its own contribution or from how it works together with related elements like a function signature. These are part of our future agenda.

\paragraph{\scalebox{0.8}{\faUser} \xspace \textbf{\textit{Practitioners.}}}
One immediate benefit of adopting \approach from the practitioners' point of view lies in its ability to provide a streamlined explanation that shows which parts of a prompt (or code parts) most strongly influenced the model's output. This, enables developers to better understand, validate, and refine LLM-generated content. As confirmed by our user study, such targeted attribution facilitates the extraction of actionable insights, thereby supporting informed decision-making throughout different stages of the software development life-cycle.

Another implication pertains to prompt optimization and emerges directly from our empirical observations. Since \approach provides fine-grained attribution scores for each semantic input feature, it enables precise refinement of the components that most significantly influence the model's behavior. For example, in a code generation setting with a prompt composed of five distinct features, \approach might reveal that just two of them (\eg $f_1$ and $f_3$) contribute over 95\% of the total attribution. If these dominant features are found to be poorly phrased or overly verbose, developers can revise them to improve clarity and reduce noise. This targeted revision is valuable in light of recent findings showing that verbose prompts can dilute critical signals and impair model performance \cite{levy2024same}.

Beyond performance gains, selective refinement offers important benefits for the \textit{sustainability} of LLM-based workflows. By systematically identifying and eliminating irrelevant or low-impact features, prompts can be compressed significantly while preserving task accuracy. This compression reduces both the cognitive burden and computational overhead of prompt engineering, yielding concrete savings in inference time and API costs, particularly when deployed at scale.

Finally, practitioners can define features that match their own conventions obtaining explanations tailored to singular developer preferences.
For example, a developer could manually split a prompt into their own manually-defined features by enforcing the conceptual structure they apply when reasoning about a task.
One developer might prefer a more general aggregation such as treating all initial assumptions as ``setup logic'', the main requirements as ``core logic'', and all special conditions as ``edge-case rules''. Another developer might instead prefer a more fine-grained view, further separating preconditions from parameter constraints, or splitting edge-case handling into individual subrules. Over time, such manually crafted features can be leveraged to train a personalized splitter, enabling \approach to produce explanations more aligned with that singular developer's reasoning style.

\paragraph{\scalebox{0.8}{\faLaptopCode} \xspace \textbf{\textit{Tool Builders.}}}
For tool builders, \approach presents a practical opportunity to integrate explainability into development workflows. Our experiments show that competitive attribution quality can be achieved even with low sampling ratio, enabling efficient real-time explanations. This suggests a straightforward IDE integration pattern: default to low-cost, low-sampling-ratio attributions for instant feedback during prompt development.
This approach makes feature-level explainability computationally feasible as an always-available development aid rather than a resource-intensive post-hoc analysis, allowing developers to continuously monitor and refine feature contributions throughout the prompt engineering process.

\section{Threats to Validity}
\label{sec:threats}

\textbf{Threats to \emph{construct validity}} concern the extent to which our experimental setup accurately captures the theoretical constructs under investigation--in this case, whether \approach meaningfully explains the behavior of LLMs in software engineering tasks, particularly code generation and summarization.
\approach, as a method grounded in SHAP, produces explanations that are inherently \emph{local} and instance-specific because Shapley values are computed relative to the set of co-occurring features in a given instance. Contextual variations--such as changes in prompt structure, feature order, or the inclusion of additional semantic elements—can shift the resulting importance distribution.
A key threat lies in how we define semantic input features. For code generation, we group tokens into components such as the docstring summary, metadata sections, and function signatures. For code summarization, we segment the code into higher-level blocks (\eg AST nodes like function definitions and their statements). Although these abstractions are grounded in developer practice, language models may instead rely on statistical cues, subword artifacts, or positional encodings that do not align with our chosen boundaries. Thus, there is a risk that our abstractions capture only correlations rather than the mechanisms that actually drive the model predictions. Future work should investigate whether feature definitions genuinely reflect what models rely on or merely approximate them.

A second threat arises from our choice of comparator metrics. In code generation, we use CodeBLEU to assess output similarity, which, despite being tailored for code, may not fully capture behavioral equivalence. For summarization, we rely on BERTScore to measure semantic similarity between summaries. Like other text-based metrics, it may not recognize valid paraphrases or reward superficial overlap. Although our evaluation methods are standard and practical, they may not fully align with the intuition of the developer or with the true functional correctness.

\textbf{Threats to \emph{internal validity}} pertain to methodological choices that may have affected our results.
In our study, attribution variance stemming from the inherent non-determinism of LLMs is only partially mitigated, as we relied on greedy decoding for all conditions to keep the experimental procedure tractable.

\textbf{Threats to \emph{external validity}} concern the generalizability of our findings beyond the specific setting of our experiments. 
First, our analysis is grounded in a specific benchmark of paired prompts derived from the \texttt{BigCodeBench} and \texttt{CodeSearchNet} benchmarks, which, although widely adopted, may not represent the full spectrum of real-world code generation or summarization tasks.
Second, we focus on a single programming languages (\ie Python), and additional work is needed to extend our approach to other languages, such as Java, C++, \etc
Third, we evaluated two representative model families \texttt{Qwen2.5-Coder} (3B/7B/32B) and \texttt{GPT-4.1-nano}, spanning open-weights and closed-source settings; although these are widely used and representative of the state-of-the-art, extending the analysis to other models would strengthen external validity.

\textbf{Threats to \emph{conclusion validity}} refer to the correctness of inferences drawn from our experimental data. We report descriptive comparisons across models, datasets, and baselines, but we do not establish causal relationships beyond noise-injection validation.
However, to mitigate these threats, we employed statistical procedures (Wilcoxon signed-rank test, Cliff's $\delta$) to compare attribution distributions, and Holm's correction for multiple comparisons.
Our assessments also rely on the perception of the practitioner rather than ground-truth explanations of the behavior of the model. Moreover, multi-collinearity among correlated code features (\eg docstrings, signatures, and implementations) can distort Shapley attributions by artificially distributing importance across dependent elements. Although \approach provides interpretable results, we cannot claim that these align with the actual internal decision-making processes of the LLMs \cite{cao2025systematic}.

\section{Conclusions and Future Work}
\label{sec:conclusions}

In this work, we introduced \approach, the first fully automated black-box explainability framework explicitly designed for software engineering tasks, particularly generative bi-modal tasks, including code generation and code summarization. By attributing model outputs to semantically meaningful features, \approach narrows the gap between opaque model behavior and the way developers naturally reason about code. Our empirical evaluation shows that \approach not only outperforms random baselines and LLM-as-an-attributor to identify irrelevant inputs, but also produces explanations that align closely with the intuition of practitioners. These findings position \approach at the forefront of explainable software engineering automation, laying the groundwork for future efforts in the area that should focus and examine the statistical rigor of attribution methods, incorporate broader evaluation metrics beyond similarity-based scores, and address the challenges of aligning explanations with human reasoning and expectations.

\section{Acknowledgments}
This publication is part of the project PNRR-NGEU which has received funding from the MUR – DM 118/2023.
This work has been partially supported by the European Union - NextGenerationEU through the Italian Ministry of University and Research, Projects PRIN 2022 ``DevProDev: Profiling Software Developers for Developer-Centered Recommender Systems'', grant n. 2022S49T4W, CUP: H53D23003610001.
This work has been partially supported by the European Union-NextGenerationEU through the Italian Ministry of University and Research, Projects PRIN 2022 “QualAI: Continuous Quality Improvement of AI-Based Systems,” Grant no. 2022B3BP5S, CUP: H53D23003510006.

\newpage

\bibliographystyle{ACM-Reference-Format}
\bibliography{main}

\end{document}